\documentstyle[12pt,epsf]{article}
\newcommand{\bi}{\bibitem}
\newcommand{\be}{\begin{eqnarray}}
\newcommand{\ee}{\end{eqnarray}}
\newcommand{\nn}{\nonumber}
\catcode`\@=11
\def\lsim{\mathrel{\mathpalette\@versim<}}
\def\gsim{\mathrel{\mathpalette\@versim>}}
\def\@versim#1#2{\vcenter{\offinterlineskip
\ialign{$\m@th#1\hfil##\hfil$\crcr#2\crcr\sim\crcr } }}
\catcode`\@=12

\begin{document}
\pagestyle{empty}
\begin{flushright}
KANAZAWA-99-23\\
October 1999
\end{flushright}

\begin{center}
{\Large\bf  Kaluza-Klein Thresholds \\ 
and \\ 
Regularization (In)dependence}
\end{center} 

\vspace{0.5cm}
\renewcommand{\thefootnote}{\alph{footnote}}
\begin{center}
{\sc Jisuke Kubo}, 
{\sc Haruhiko Terao} and
{\sc George Zoupanos}
\footnote{Permanent address:
Physics Department, Nat. Technical University, 
GR-157 80, Zografou, Athens, Greece.
Partially supported by the E.C. projects, 
ERBFMRXCT960090 and   ERBIC17CT983091.
}
\end{center}
\begin{center}
{\em 
Institute for Theoretical Physics, 
Kanazawa  University, 
Kanazawa 920-1192, Japan}
\end{center}

\vspace{1cm}
\begin{center}
{\sc\large Abstract}
\end{center}

\noindent
We present a method to control the regularization scheme dependence
in the running of couplings in Kaluza-Klein theories.
Specifically we consider the scalar theory in five 
dimensions, assuming that one dimension is compactified and we study
various regularization schemes in order to analyze concretely the
regularization scheme dependence of the Kaluza-Klein threshold
effects. We find that  in  one-loop order, although
the $\beta$-functions are different for the different schemes,
the net difference in the running of the coupling
among the different schemes is very small for the entire range
of energies. Our results have been extended to include more 
than one radii, and the gauge coupling unification is re-examined. 
Strings are also used as a regulator. We obtain a particular 
regularization scheme of the effective field theory which
can accurately describe the string Kaluza-Klein threshold effects.
 

\newpage
\pagestyle{plain}
\pagenumbering{arabic}
\renewcommand{\thefootnote}{\arabic{footnote}}
\setcounter{footnote}{0}

\section{Introduction}
Recently there have been renewed interests in Kaluza-Klein theories
with a large compactification radius 
\cite{antoniadis1}--\cite{phenomenology}.
The radius can be so large that quantum effects, in particular the 
running of coupling constants in these theories, can drastically change 
the traditional  picture of unification of the gauge couplings 
\cite{dienes1}.
These higher dimensional field theories might
appear as the low-energy effective
theory of a string theory with certain $D$-branes 
\cite{dienes1}--\cite{kakushadze1}
\footnote{See also \cite{antoniadis5} and references therein.},
and contain towers of massive Kaluza-Klein excitations.
It is indeed this tower of excitations which 
gives rise to the quantum effect that modifies the behavior
of the coupling from the logarithmic to the power law running 
\cite{veneziano,dienes1},  
thereby changing the usual unification scenario. The results of refs. 
\cite{veneziano,dienes1,kakushadze2,ross2} 
indicate, moreover, that this quantum effect can be computed to a 
certain extent within the framework of the effective 
field theory 
\footnote{See also refs.\cite{ross1}--\cite{mohapatra}.
In refs. \cite{antoniadis6,kobayashi1} quantum corrections
to supersymmetry breaking parameters also have been computed.}.
That is, physics below the string scale but above the compactification
scale (the massive string states are suppressed in this regime), 
can be well described by a field theory 
\cite{kakushadze2,ross2} 
which is unrenormalizable by power counting.
Furthermore, it is  expected  that if extra dimensions are
compactified  the unrenormalizable theory effectively appears at low 
energies  in four dimensions as a renormalizable theory, in which the 
massive Kaluza-Klein excitations are completely decoupled.
How can this decoupling become possible in unrenormalizable theories?
In this paper we are motivated by the desire to gain a deeper 
understanding on the Kaluza-Klein thresholds within the framework of 
(unrenormalizable) field theory.

A pragmatic way to define an unrenormalizable theory
is to cutoff the high energy modes by a regularization.
However, in contrast to renormalizable theories, it is no longer
obvious that  the ``physical'' quantities are truly independent of 
the regularization. Therefore, the regularization scheme dependence
must be much more serious than the renormalization scheme 
dependence in renormalizable theories.
To our knowledge there exists no complete solution to this problem.
Our task in this paper is to take a step in solving this
important problem in the case of  the running of couplings. 
(Note that the running of couplings is strongly dependent on
regularization.) 
To simplify the situation we will first consider the Kaluza-Klein theory
for  the scalar field  with a self-interacting coupling in 
five dimensions where we assume that one extra dimension is
compactified on a circle with radius $R$.  
We will assume that the leading high energy behavior
in the running of the coupling is fixed by its canonical dimension
without any corrections.
From this we postulate in section 2 the form of the $\beta$-function
of the dimensionally-reduced, four-dimensional
theory containing the  Kaluza-Klein tower and show that the
regularization scheme dependence in the running of the coupling
can be systematically controlled in the asymptotic regime 
(i.e. in energy scales much higher than the compactification scale $1/R$).
In other words, different regularization schemes in the asymptotic 
regime can be related  by a ``finite'' transformation of the 
corresponding coupling,
in a very similar way as in the case of a renormalizable theory. 
Consequently, the net difference among regularization schemes
can originate only from the threshold effect of the first Kaluza-Klein 
excitations. 

In  section 3 we will consider various regularization schemes,
the Exact renormalization group (ERG) scheme 
\cite{wilson}--\cite{legendre} 
\footnote{The Exact RG scheme has been applied to
the Kaluza-Klein theories in refs. \cite{kobayashi1,clark}.},
the momentum subtraction scheme and the proper time regularization
scheme \cite{dienes1}, and we introduce the notion of effective dimension
(which has been used to absorb the environmental effects at finite
temperature into a redefinition of the spatial dimension \cite{connor})
to absorb the Kaluza-Klein threshold effects into the effective dimension.
Studying the effective dimension as a function of energy, we will observe
the smooth transition of the coupling from the logarithmic behavior to
the power law behavior, which shows that the effective dimension contains
the full information of the Kaluza-Klein threshold effects.
In section 4 we will then investigate the one-loop evolution
of the coupling in these schemes as well as in the one-$\theta$ function 
and the successive-$\theta$ functions approximation schemes, 
and extract the net difference among the
schemes by taking into account appropriately chosen finite 
transformations of the coupling.
We will find that the net difference in the evolution of the coupling
among the different regularization schemes is very small.
At first sight this looks like a surprising result, but in fact it
only generalizes the result of  ref.~\cite{dienes1} that the 
one-$\theta$-function approximation is a very good approximation to 
the proper time regularization scheme.
We expect that this feature of the threshold effect, which does not 
exist in the usual renormalizable theories, is very general
in Kaluza-Klein theories, and therefore our result obtained in a
scalar field theory can be extended to gauge theories.
In fact we examine gauge coupling unification
of the minimal supersymmetric standard model (MSSM)
with the Kaluza-Klein tower only in the gauge and Higgs 
supermultiplets (proposed in ref.~\cite{dienes1})
by taking into account the Kaluza-Klein threshold effect.
We find that because of a compensating
mechanism which exists in this model the prediction 
of $\alpha_3(M_Z)$ does not change practically.

Finally we compare the Kaluza-Klein threshold effect
in a string theory and its effective field theory.
We will observe that  string theory result
averages those of field theory in different
regularizations  with a definite weight,
and that it defines an average 
regularization scheme of the effective theory.
Remarkably, we will find that 
the effective theory
with the average  regularization scheme
can accurately describe the Kaluza-Klein threshold effects of 
the string theory.

\section{Controlling regularization scheme dependence}

In order to concentrate on the regularization scheme dependence and 
to avoid problems that might complicate our task
(e.g. violation of gauge symmetries through regularization),
we consider the Kaluza-Klein  theory for the scalar field $\phi$  in 
Euclidean five dimensions where we assume that the one extra dimension
is compactified on a circle with  radius $R$.  
We denote the extra coordinate by $y$ and the four dimensional
coordinates by $x_{\mu}$. 
The starting five dimensional action is
\be
S^{(5)} &=& \int_{0}^{2\pi R} d y 
\int d^4x \left\{~
\frac{1}{2}(\partial_{y}\tilde{\phi})^2+
\frac{1}{2}(\partial_{\mu}\tilde{\phi})^2+
\frac{\tilde{m}^2}{2}\tilde{\phi}^2+\frac{\tilde{\lambda}}{4!}
\tilde{\phi}^4~\right\}~. 
\label{action-d}
\ee
The scalar field satisfies the boundary condition
\be
\tilde{\phi}(x,y) &=&\tilde{\phi}(x,y+2\pi R) ~,
\label{bcs}
\ee
which implies that the field can be expanded as
\be
\tilde{\phi}(x,y) &=&
\sum_{ n \in {\bf Z}}~\tilde{\phi}_{ n}(x)~\exp (i \omega_{n} y)
~,~~~~\omega_n = \frac{n}{R}~,\\
(\tilde{\phi}_{ n})^{*} &= &\tilde{\phi}_{- n}~,\nn
\label{expansion1}
\ee
where the last equality follows from the reality of $\tilde{\phi}$.
To define the four dimensional action we redefine
the field and the parameters according to
\be
\tilde{\phi}_{ n} &\to &  \phi_n=\tilde{\phi}_{ n}/\sqrt{2\pi R}~,~
\tilde{\lambda} \to \lambda=2\pi R \tilde{\lambda}~,
~\tilde{m}^2 \to m^2 = \tilde{m}^2~.
\label{scaling2}
\ee
We then obtain
\be
S=\int d^4 x \left\{
\frac{1}{2}\sum_{ n \in {\bf Z}}\phi_{-n}[
-\partial_{\mu}^2+m^2+\omega_{n}^2]\phi_{n}
+\frac{\lambda}{4!} \sum_{n_i \in {\bf Z}}
\delta_{n_1+n_2 + n_3+ n_4,0}
\phi_{ n_1} \phi_{ n_2}\phi_{ n_3}\phi_{ n_4} \right\}.
\label{action4}
\ee

The canonical dimension of the original coupling $\tilde{\lambda}$, 
defined in (\ref{action-d}), is $-1$, and so one expects that 
$\tilde{\lambda}$ behaves like $\tilde{\lambda}\sim \Lambda^{-1}$
for large $\Lambda$. 
As we have announced in the introduction, we assume that
the leading order behavior does not suffer from any scaling
violation. 
This implies that the $\beta$-function of the coupling
$\tilde{\lambda}$ in the five dimensional theory can be written as
\be
\tilde{\beta} & =&\Lambda\frac{d\tilde{\lambda}}{d \Lambda}
\simeq \sum_{n\geq 2} ~
 \tilde{b}_{n} \tilde{\lambda}^n \Lambda^{n-1}~
\label{beta2}
\ee
for large $\Lambda$.
In general it is expected that the expansion coefficients 
$\tilde{b}_{n}$ depend on the regularization scheme employed, 
even in the lowest order, i.e. $\tilde{b}_2$.
Next if the four dimensional theory defined by (\ref{action4}) should 
be related or be an approximation to the five dimensional theory, then 
the $\beta$-function of $\lambda$ should approach, upon the rescaling
(\ref{scaling2}), the form (\ref{beta2})
of the five dimensional theory in the large $\Lambda$ limit.
From this consideration we postulate that the $\beta$-function of 
$\lambda$ for large $\Lambda$ has the form
\be
\beta = \Lambda \frac{d \lambda}{d \Lambda}
= \sum_{n\geq 2} ~
b_{n} \lambda^n (R\Lambda)^{n-1} + \Delta\beta~,
\label{beta3}
\ee
where $\Delta\beta$ stands for sub-leading terms which vanish in the 
$\Lambda \to \infty$ limit (with $\lambda$ kept constant).
Like the coefficients $\tilde{b}_n$, the expansion coefficients $b_{n}$
depend on the regularization scheme.

Before we proceed, let us note the following remark.
In usual renormalizable gauge theories, the renormalization scheme
dependence has been systematically studied in ref.~\cite{stevenson}.
We recall that in this case there exists  not only  the complete
parametrization of the renormalization scheme dependence, but also 
there exist exact, renormalization-scheme independent quantities
at a given order in perturbation theory.
It might be possible to generalize these results to
non-renormalizable theories.
However, the generalization will not be straightforward, because in 
renormalizable theories there exist quantities 
(i.e. the physical quantities) that are formally ensured
to be independent of renormalization scheme if
one considers all orders of the perturbative expansions.

Given the asymptotic form of the $\beta$-function 
(\ref{beta3}) in four dimensions, we analyze the regularization 
dependence, and consider the change of the scheme, 
$\lambda \to \lambda'$.
We require that this transformation satisfies:
\be
~\lambda' &\sim& 1/\Lambda ~~\mbox{as}~~
\Lambda \to \infty~.
\label{require1}
\ee
It is now straightforward to see that the most general form of the
transformation that keeps the asymptotic form (\ref{beta3}) invariant
can be written as
\be
\lambda & \to & \lambda' = \sum_{n \geq 1}
c_{n} \lambda^n (R\Lambda)^{n-1}+\Delta_{\lambda}~,
\label{transform1}
\ee
where $c_{1}=1$, and $\Delta_{\lambda}$ 
stands for the terms in the sub-leading orders.
It may be worthwhile to see how the transformation
(\ref{transform1}) changes the coefficients $b_n$ in lower orders.

To this end, we consider up to the next-to-leading order:
\be
\beta &=& b_2 \lambda^2(R\Lambda)
+b_3 \lambda^3 (R\Lambda)^2+\dots~,\\
\lambda' &=& \lambda (1+ c_2 \lambda \Lambda 
+c_3 \lambda\Lambda^2+\dots)~,
 \label{beta4}
\ee
and find that
\be
\beta' &=&\Lambda \frac{d \lambda'}{d \Lambda}
= b_{2}^{\prime} \lambda^{\prime 2}(R\Lambda)
+b_{3}^{\prime} \lambda^{\prime 3} (R\Lambda)^2+\dots~,\\
\mbox{with} & &  b_{2}^{\prime}=b_2+c_2    ~,~
b_{3}^{\prime}=b_3-2c_2^2+2 c_3 ~.
\label{beta5}
\ee
So the lowest order coefficient $b_2$
has been changed, which should be contrasted to the renormalizable case
where the lowest order coefficient does not change
under the change of scheme.

We will see that the discussion above plays an important role in the 
next section in taking into account the threshold corrections of the 
Kaluza-Klein modes and comparing the threshold effects obtained in 
different regularization schemes with each other.

\section{Explicit calculations in different schemes}
In order to describe the Kaluza-Klein threshold effects and obtain
the power-law behavior of the effective couplings, it is important to 
treat the decoupling of the massive Kaluza-Klein modes faithfully. 
In this section we adopt three different regularization schemes 
of this kind;
1) the Exact RG (ERG) scheme (the Legendre flow equations)
\cite{legendre},
2) the momentum subtraction (MOM) scheme and 
3) the proper time regularization (PT) scheme
\cite{dienes1}.
In what follows,
we will study the regularization scheme dependence in the 
$\beta$-function and compare the results of these schemes.

\subsection{Exact RG approach}

\noindent
{\bf A.} {\em Uncompactified case}

\noindent 
The theory defined by (\ref{action-d}) is (perturbatively) 
unrenormalizable by power-counting. Therefore we need to define
the theory as a cutoff theory.  
The natural framework to study low-energy physics of the 
cutoff theory is provided by the continuous Wilson 
renormalization group (RG)
\cite{wilson}--\cite{legendre}.
To illustrate the basic idea of the Wilson RG approach,
we assume for a while that the extra dimension is not compactified.
First we divide the field in the momentum space, $\phi(p)$, into
low and high energy modes according to
\be
\phi (p)=\theta(|p|-\Lambda)\phi_{>}(p)+
(1-\theta (|p|-\Lambda))\phi_{<}(p),
\label{modes}
\ee
where $\theta(|p|-\Lambda)$ denotes a proper infrared cutoff
function.
The Wilsonian effective action is defined by 
integrating out the high energy modes in the path integral; 
\be
S_{\rm eff} [~\phi_{<},\Lambda~]
=-\ln \left\{
\int {\cal D}\phi_{>} ~e^{-S[\phi_{>},\phi_{<}]}
\right\}.
\label{wilson}
\ee
It was shown in ref.~\cite{wilson,wegner} that the path integral
corresponding to the difference
\be
\delta S_{\rm eff} =
S_{\rm eff} [~\phi_{<},\Lambda+\delta\Lambda~] - S_{\rm eff} 
[~\phi_{<},\Lambda~]
\ee
for an infinitesimal $\delta\Lambda$ can be exactly carried out. 
That is, it is possible to write down a concrete expression
for the RG equation of the effective action in the form
\be
\Lambda \frac{\partial S_{\rm eff}}{\partial \Lambda}
&=& {\cal O}(S_{\rm eff})~,
\label{nrg1}
\ee
where ${\cal O}$ is a non-linear operator acting on the functional 
$S_{\rm eff}$. 
Since $S_{\rm eff}$ is a (non-local) functional of fields, one can 
think of the  RG equation (\ref{nrg1}) as coupled differential
equations for  infinitely many couplings in the effective action.
The crucial point is that ${\cal O}$ can be exactly derived for a
given theory, in contrast to the perturbative RG approach where the 
RG equations are known only up to a certain order in perturbation 
theory. This provides us with possibilities to use approximation
methods that go beyond the conventional perturbation theory 
\cite{wetterich,morris,aoki}.

There exist various (equivalent) formulations 
\cite{wegner}--\cite{legendre}
of the ERG approach, and depending on the formulation and also on the
cutoff scheme one has to consider different 
forms of the operator ${\cal O}$ 
\footnote{
The formal relations between these formulations are 
investigated in refs. \cite{scheme}.
}. 
Of the different formulations, the so-called Legendre flow equation
\cite{legendre} is the most suitable one to incorporate the decoupling 
of heavy particles
\footnote{
The Polchinski formulation \cite{polchinski} is not 
appropriate to take into account the heavy particle decoupling, 
since a cutoff is introduced  in the massless propagator.
The Wegner-Houghton equation with a sharp cutoff \cite{wegner} 
is not appropriate either because it becomes 
singular at the thresholds of the heavy particles.
}.
Here let us first briefly outline the basic
idea to derive the Legendre flow equation that
describes the evolution of the cutoff effective action.
As a first step we define the generating functional $W[J]$ of the connected 
Green's functions with an {\em infrared cutoff} $\Lambda$ by
\be
W[J] = \ln \int {\cal D}\phi\,\exp 
\left\{ ~\int\frac{d^D p}{(2\pi)^D} 
\,[\,-\frac{1}{2}\phi(-p) C^{-1}(p,\Lambda) \phi(p)]
-S_{\Lambda_{0}}[\phi]+\phi\cdot J ~\right\}~,
\label{wj1}
\ee
where $D=5$ in the present case, and 
\be
\phi\cdot J ~= ~\int\frac{d^D p}{(2\pi)^D} ~J(-p)\phi(p)~.
\ee
Here $ S_{\Lambda_{0}} $ stands for the bare action,
and  $C^{-1}(p,\Lambda)$ is a smooth cutoff function of $p^2$ and 
$\Lambda^2$ which cutoffs the propagation of the lower energy modes 
with $|p| \ll \Lambda$.  
The explicit form of the cutoff function will be specified later on.
An ultraviolet cutoff is assumed in the r.h.side of (\ref{wj1}), 
but the results we will obtain are ultraviolet finite.
One derives the RG equation for $W[J]$ for the present system by 
differentiating the both sides of eq.~(\ref{wj1}) with respect to 
$\Lambda$:
\be
\Lambda\frac{\partial W}{\partial \Lambda}
&=&-\frac{1}{2}
\int\frac{d^D p}{(2\pi)^D}
\,\left[~  
\frac{\delta W}{\delta J(p)}\, 
(\Lambda\frac{\partial C^{-1}(p,\Lambda)}{\partial \Lambda})\,
\frac{\delta W}{\delta J(-p)} \right.\nn\\
& &+\left.
(\Lambda\frac{\partial C^{-1}(p,\Lambda)}{\partial \Lambda})\,
\frac{\delta^2 W}{\delta J(p) \delta J(-p)}\,\right]~.
\label{e-p-eq}
\ee
This flow equation can be translated to that of 
the infrared cutoff effective action which is defined by Legendre
transformation:
\be
\Gamma_{\rm eff}[\phi] &=&-W[J]+J\cdot \phi
 -\frac{1}{2}\int\frac{d^D p}{(2\pi)^D} 
\,\phi(-p) C^{-1}(p,\Lambda) \phi(p)~,
\label{gamma1}
\ee
yielding 
\be
\Lambda\frac{\partial \Gamma_{\rm eff}}{\partial \Lambda} 
=
\frac{1}{2}\,\int\frac{d^D p}{(2\pi)^D}\,
\Lambda\frac{\partial 
C^{-1}(p,\Lambda)}{\partial \Lambda}~
\left[~
C^{-1}(p,\Lambda)+
\frac{\delta^2\Gamma_{\rm eff}}{\delta \phi(p)\delta \phi(-p)}   
~\right]^{-1}~.
\label{rge1}
\ee

As we see from the above RG equations (\ref{e-p-eq}) and (\ref{rge1}), 
they explicitly depend on the cutoff function $C(p,\Lambda)$. 
The physical quantities  should be independent of it.  
However, this is not automatically
ensured as we have emphasized in section 1.
In this paper we use the smooth
cutoff function  \footnote{
This cutoff function has been used
in order to calculate critical exponents in lower dimensions
\cite{morris,aoki}.
}
\be
C(p,\Lambda) = \frac{1}{\Lambda^2}~
\left(\frac{p^2}{\Lambda^2}\right)^k~,
\label{cutoff}
\ee
where $k$ is a positive integer satisfying $k > D/2-2$.
To see that this is in fact an infrared cutoff, we look at
the (massless) propagator which can be read off from the
generating  function (\ref{wj1}):
\be
\frac{1}{p^2}\theta_{\Lambda}(p^2/\Lambda^2)~~~
\mbox{with} ~~
\theta_{\Lambda}(s)= \frac{s^{k+1}}{1+s^{k+1}}~.
\label{theta}
\ee
One sees that the cutoff profile becomes sharper for bigger $k$. 
(Later we will examine  the cases of $k=1$ (slow cutoff) and $k=5$ 
(sharp cutoff) .)
For the cutoff function (\ref{cutoff}) the ERG equation (\ref{rge1}) 
becomes finally
\be
\Lambda\frac{\partial \Gamma_{\rm eff}}{\partial \Lambda} &=&
(k+1)\,\int\frac{d^D p}{(2\pi)^D}\,
\left[~
1+\frac{1}{\Lambda^2}~ (\frac{p^2}{\Lambda^2})^k~
\frac{\delta^2\Gamma_{\rm eff}}{\delta \phi(p)\delta \phi(-p)}   
~\right]^{-1}~.
\label{rge2}
\ee

The RG equations (\ref{rge1}) and (\ref{rge2}) are
integro-differential equations of first order in $\Lambda$
defined on a space of functionals of fields.
In general, these RG equations cannot be solved exactly, and therefore
one has to apply some approximation method.
Clearly, assumptions which are made for a specific approximation 
method should be confirmed in a case-by-case analysis
at a desired level within the approximation scheme. 
In the derivative expansion approximation \cite{wetterich,morris},
one assumes that the Legendre effective action $\Gamma_{\rm eff}[\phi]$ 
can be written as a space-time integral of a (quasi) local function
of $\phi$, i.e.,
\be
\Gamma_{\rm eff}[\phi] &=&
\int d^D x\,(\,\frac{1}{2}\,
\partial_{\mu}\phi\partial_{\mu}\phi\,Z_{\phi}(\phi)
+V(\phi,\Lambda)+\dots~)~,
\label{ansatz1}
\ee
where $\dots$ stands for terms with higher order derivatives
with respect to the space-time coordinates.
In the lowest order of the derivative expansion (the local potential
approximation (LPA) \cite{LPA}), 
there is no wave function renormalization 
($Z_{\phi}(\phi)=1$) and the RG equation for the effective potential
$V$ can be obtained. 
It is more convenient to work with the RG equation for
dimensionless quantities, which makes the  scaling properties
more transparent. 
To this end, we scale the quantities according to
\be
p &\to \Lambda p~,~\phi \to \Lambda^{D/2-1} \phi~,~
V \to \Lambda^{D} V~.
\label{scaling1}
\ee
Then the RG equation is found to be
\be
\Lambda\frac{\partial V}{\partial \Lambda}
& =&-D V+\frac{D-2}{2} \phi V'
+(k+1) \frac{A_D}{2} \int_{0}^{\infty} d s\,s^{D/2-1}\,
\left[~
 1+s^{k} (s+V^{\prime\prime})~
\right]^{-1}~,
\label{rge3}
\ee
where the prime on $V$ stands for the derivative with respect to 
$\phi$, and
\be
A_D &=& \frac{2^{1-D}}{\pi^{D/2} \Gamma(D/2)} 
\label{ad}
\ee
is the $D$ dimensional angular integral.

Since we are not interested in the non-perturbative calculations,
but in the threshold effects of the Kaluza-Klein modes in the weak
coupling regime, we assume that
the potential $V$ in eq.~(\ref{rge3}) can be expanded as
\be
V &=& \frac{1}{2}m^2 \phi^2 +\frac{1}{4!}\lambda \phi^4
+\dots~.
\label{lpa1}
\ee
This gives the RG equations for the effective couplings;
\be
\Lambda \frac{\partial m^2}{\partial \Lambda} &=& 
-2 m^2 -\lambda (k+1) \frac{A_D}{2}
\int_{0}^{\infty}d s s^{D/2-1} 
\frac{s^k}{(1+s^{k+1}+m^2 s^k)^2}~,
\label{rge-m1}\\
\Lambda \frac{\partial \lambda}{\partial \Lambda} &=& 
(D-4) \lambda +6 \lambda^2 (k+1) \frac{A_D}{2}
\int_{0}^{\infty}d s s^{D/2-1}
 \frac{s^2k}{(1+s^{k+1}+m^2 s^k)^3}~.
\label{rge-l1}
\ee

It is worth-considering the massless case, because
 $m^2$ (which is the mass of the zero mode in Kaluza-Klein 
theories) is
small as compared 
to the scale of compactification in most practical situations.
Then we find that
\be
\Lambda \frac{\partial \lambda}{\partial \Lambda} &=& 
(D-4) \lambda+\frac{3}{2} A_D B(D,k) \lambda^2~,
\label{rge-l21}\\
B(D,k) &=& 2 (k+1) \int_0^{\infty}d s s^{D/2-1} 
\frac{s^{2k}}{(1+s^{k+1})^3}~.
\label{bd}
\ee
In terms of the original, dimensional coupling 
(i.e. $\lambda \to \Lambda^{D-4}\lambda$),
the evolution equation (\ref{rge-l21}) becomes
\be
\Lambda \frac{\partial \lambda}{\partial \Lambda} &=& 
\frac{3}{2} A_D B(D,k) \Lambda^{4-D}\lambda^2~,
\label{rge-l31}
\ee
recovering the power law behavior of the scalar coupling.
Note that the constant $B(D,k)$ depends explicitly
on  $k$ and $D$ ($k > D/2-2$). We find, for instance, that
\be
& &B(4,k)=1~~~\mbox{(for all $k$)},\nn \\
& &B(5,1)=1.3884\cdots,~~~B(5,2)=1.22173\dots,~~~B(5,5)=1.09581\dots~,\nn\\
& &B(6,5)=1.22173\cdots~.
\label{bss}
\ee
As we can see from eq.~(\ref{rge-l31}), 
the $k$ dependence (regularization dependence) can be
absorbed into the redefinition of the coupling $\lambda$ for $D \neq 4$,
according to the discussion of section 2.
Fortunately, at $D=4$ (where the power behavior disappears)
the $k$ dependence vanishes, as we see from (\ref{bss}).

\vspace{0.7cm}
\noindent
{\bf B.} {\em Compactified case}

\noindent
Now we would like to come to the case in which the extra spatial 
dimension is compactified on a circle with  radius $R$.  
What follows is a continuation of ref.~\cite{kobayashi1} and 
a justification of the main assumptions that have been made there. 
The results of ref.~\cite{kobayashi1} were also
based on the recent developments that have been made concerning the
renormalization properties of the supersymmetry breaking parameters in
supersymmetric gauge theories \cite{susybreak}.

The starting four dimensional action is given by eq.~(\ref{action4}).
To define the generating functional $W[J]$, we introduce  
like eq.~(\ref{wj1}) the infrared cutoff term
\footnote{
In the $D=5$ case we may adopt the cutoff functions that
cutoff only the four dimensional momentum, since the infinite sum 
of the Kaluza-Klein modes converges.
Here we have adopted a cutoff function that can
be applied to the 
$D>5$ cases as well. 
It is worth-noting that in finite temperature 
field theory \cite{temperature} one  obtains
Legendre flow equations, which are similar to eq.~(\ref{rge5}).
}
\be
\Delta S &=& \frac{1}{2} \int_0^{2\pi R} d y \int d^4 x \phi(x,y)
C^{-1}(-i\partial_x,-i\partial_y,\Lambda)\phi(x,y)\nn\\
& =&
\frac{1}{2}\int \frac{d p^4}{(2\pi)^4} \sum_{n}\phi_{-n}(-p)
C^{-1}(p,\omega_{n},\Lambda)\phi_{n}(p)~.
\label{cutoff3}
\ee
Then we follow exactly the same procedure as in the uncompactified 
case, and obtain the RG equation for the effective action:
\be
\Lambda\frac{\partial \Gamma_{\rm eff}}{\partial \Lambda} 
&=& (k+1)\,\int\frac{d^4 p}{(2\pi)^4}\, \sum_{ n}~\Delta_{n,-n}(p,-p)~,
\label{rge4}\\
\Delta^{-1}_{n,- m}(p,-p) &=&
\delta_{n,-m} + \frac{1}{\Lambda^2}~
(\frac{p^2+\omega_{n}^2}{\Lambda^2})^k~
 \frac{\delta^2\Gamma_{\rm eff}}
{\delta\phi_{n}(p)\delta \phi_{-m}(-p)}~,
\ee
where we have used the infrared cutoff function
\be
C(p,\omega_{n},\Lambda) & = & \frac{1}{\Lambda^2}
 [~\frac{p^2+\omega_{n}^2}{\Lambda^2}   ~]^k~.
\label{cutoff4}
\ee
This infrared cutoff function is motivated from
that of the uncompactified case (see eq.~(\ref{cutoff})).
Calculating the propagator for a Kaluza-Klein mode,
we  find that it is multiplied by the cutoff function
\be
\theta_{\Lambda}(p,\omega_{n})
&=&\frac{[~p^2/\Lambda^2+(\omega_{n}/\Lambda)^2~]^{k+1}
}{1+[~p^2/\Lambda^2+(\omega_{n}/\Lambda)^2~]^{k+1}}~,
\label{cutoff5}
\ee
from which we see that not only high frequency modes
are  integrated out in  the effective action,
but also heavy mass modes.

We keep following  the uncompactified case.
In the lowest order of the derivative expansion 
approximation 
the Legendre effective action $\Gamma_{\rm eff}[\phi]$ 
has the same  form as (\ref{ansatz1}), i.e.,
\be
\Gamma_{\rm eff}[\phi_n] &=&
\int d^4 x\,\left(\,\frac{1}{2}\,\sum_n
\partial_{\mu}\phi_{n}\partial_{\mu}\phi_{-n}
+V(\phi_n,\Lambda)+\dots~\right)~,
\label{ansatz2}
\ee
and the potential has the same structure as the tree level
\be
V =
\frac{m^2}{2}\sum_{ n}\phi_{-n}
\left[ -\partial_{\mu}^2+m^2+\omega_{n}^2~\right]\phi_{n}
+\frac{\lambda}{4!} \sum_{n_i}
\delta_{n_1+n_2 + n_3+ n_4,0}
\phi_{ n_1} \phi_{ n_2}
\phi_{ n_3}\phi_{ n_4}+\dots~.
\label{potential2}
\ee
Then we scale the quantities
according to (\ref{scaling1}) and inserting 
the ansatz (\ref{ansatz2}) with $V$ given in eq.~(\ref{potential2})
into eq.~(\ref{rge4}) to
obtain
\be
\Lambda\frac{\partial V}{\partial \Lambda}
& =&-4 V+\sum_n\phi_n\frac{\partial V}{\partial\phi_n}\nn\\
& &+(k+1) \frac{A_4}{2} \sum_{n}
\int_{0}^{\infty} d s\,s^{4/2-1}\,
\left[~
 1+s^{k} \left(s+ \frac{\delta^2 V}
{\delta\phi_{n}(p)\delta \phi_{-n}(-p)}\right)~
\right]^{-1}~.
\label{rge5}
\ee
Expanding the potential in eq.~(\ref{rge5}) we finally obtain the RG 
equations in one-loop approximation as
\be
\Lambda \frac{\partial m^2}{\partial \Lambda} 
&=& 
-2 m^2 -\lambda (k+1) \sum_n~\frac{1}{16\pi^2}
\int_{0}^{\infty}d s s 
\frac{s_n^k}{(1+s_{n}^{k+1}+m^2 s^k)^2}~,
\label{rge-m2}\\
\Lambda \frac{\partial \lambda}{\partial \Lambda} &=& 
6 \lambda^2 (k+1) \sum_n~\frac{1}{16\pi^2}
\int_{0}^{\infty}d s s
\frac{s_{n}^{2k}}{(1+s_{n}^{k+1}+m^2 s^k)^3}~,
\label{rge-l2}
\ee
where
\be
s_n &=& s +(\frac{n}{R \Lambda})^2~~\mbox{and}~~
~\frac{A_4}{2}=\frac{1}{16\pi^2}~.
\ee
These RG equations
 should be compared with eqs.~(\ref{rge-m1}) and (\ref{rge-l1}) found
in the uncompactified case.
Note that the sum over $n$ is convergent, and
this convergence is ensured by the infrared cutoff function
$C$ given in eq.~(\ref{cutoff4}) introduced in eq.~(\ref{cutoff3}).

\subsection{Effective dimension}

Here we would like to understand how the power behavior of the
coupling for large $\Lambda$ occurs in the ERG scheme. 
To simplify things we consider the massless case.
The evolution equation (\ref{rge-l2}) in the massless limit
becomes
\be
\Lambda \frac{d \lambda}{d\Lambda} 
=\frac{3}{16\pi^2}~\epsilon_k(R\Lambda)~\lambda^2~,
\label{rge-l3}
\ee
where
\be
\epsilon_k(R\Lambda) =
2 (k+1)\sum_n~ \int_{0}^{\infty}d s s
 \frac{s_{n}^{2k}}{(1+s_{n}^{k+1})^3}~,~
s_n = s+\left(\frac{n}{R\Lambda}\right)^2~.
\label{epsilon1}
\ee
The function $\epsilon_k(R\Lambda)$ approaches to $1$ as $R\Lambda \to 0$,
(i.e. $\Lambda \to 0$), while as $ R\Lambda\to \infty$ its
derivative, $d \ln\epsilon_k /d \ln\Lambda $,  approaches 
to a constant independent of $\Lambda$.
In figs.~1 and 2 we plot $\ln \epsilon_k$ 
as a function of $t=\ln (R\Lambda)$ for $k=1$ and $5$,
respectively, where $N$ stands for  the value of $n$ in the r.h.side
of eq.~(\ref{epsilon1}) at which the sum is truncated
(i.e. $|n| \leq N$).
\begin{figure*}[p]
\epsfysize=0.4\textheight   
\centerline{\epsffile{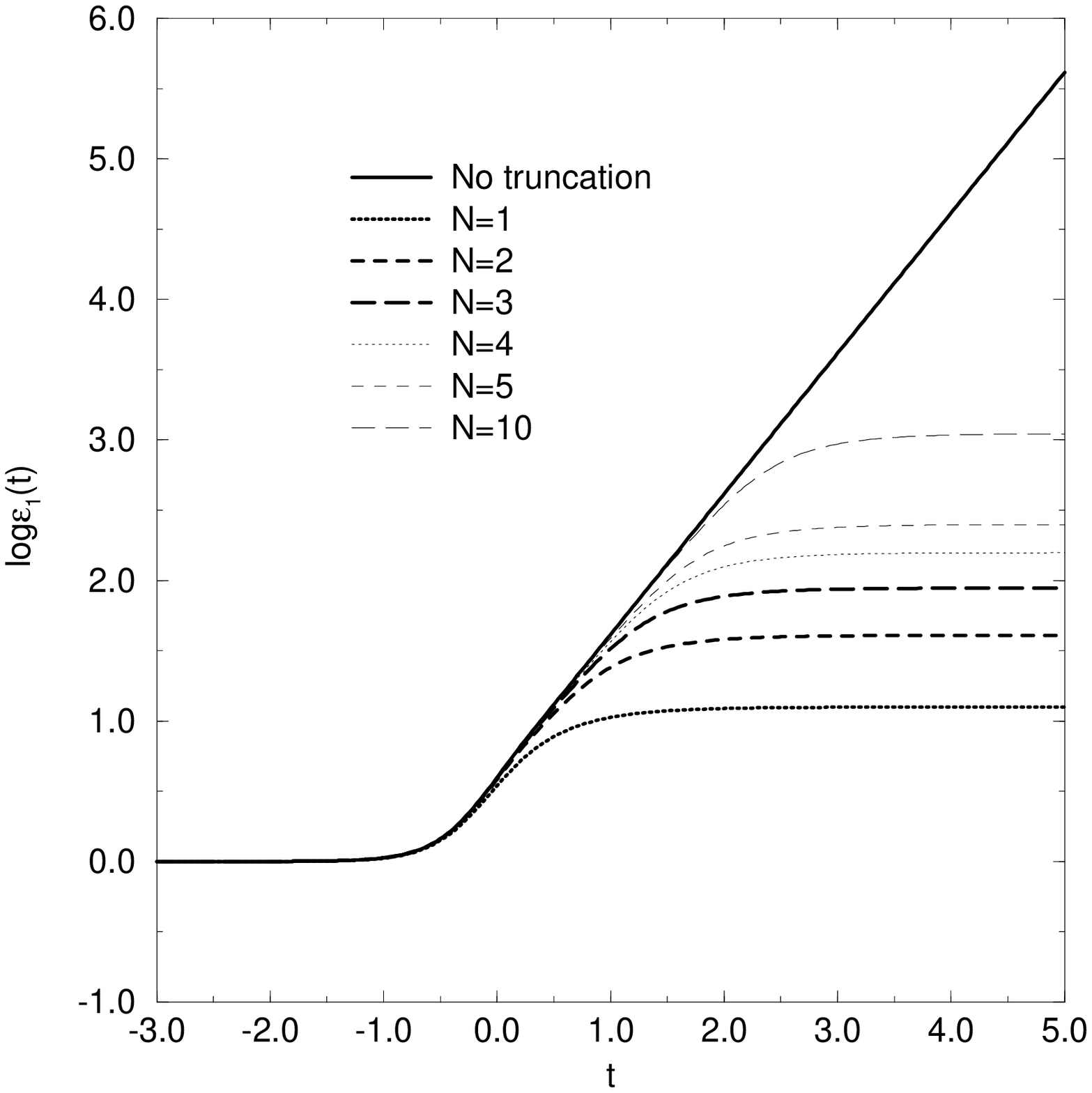}}
\label{fig:1}
\caption{
The function $\ln\epsilon_k(t)$ with $k=1$ where $t=\ln(R\Lambda)$.
$N$ stands for the cutoff in the sum over the Kaluza-Klein
excitations.
}
\vspace{5mm}
\epsfysize=0.4\textheight  
\centerline{\epsffile{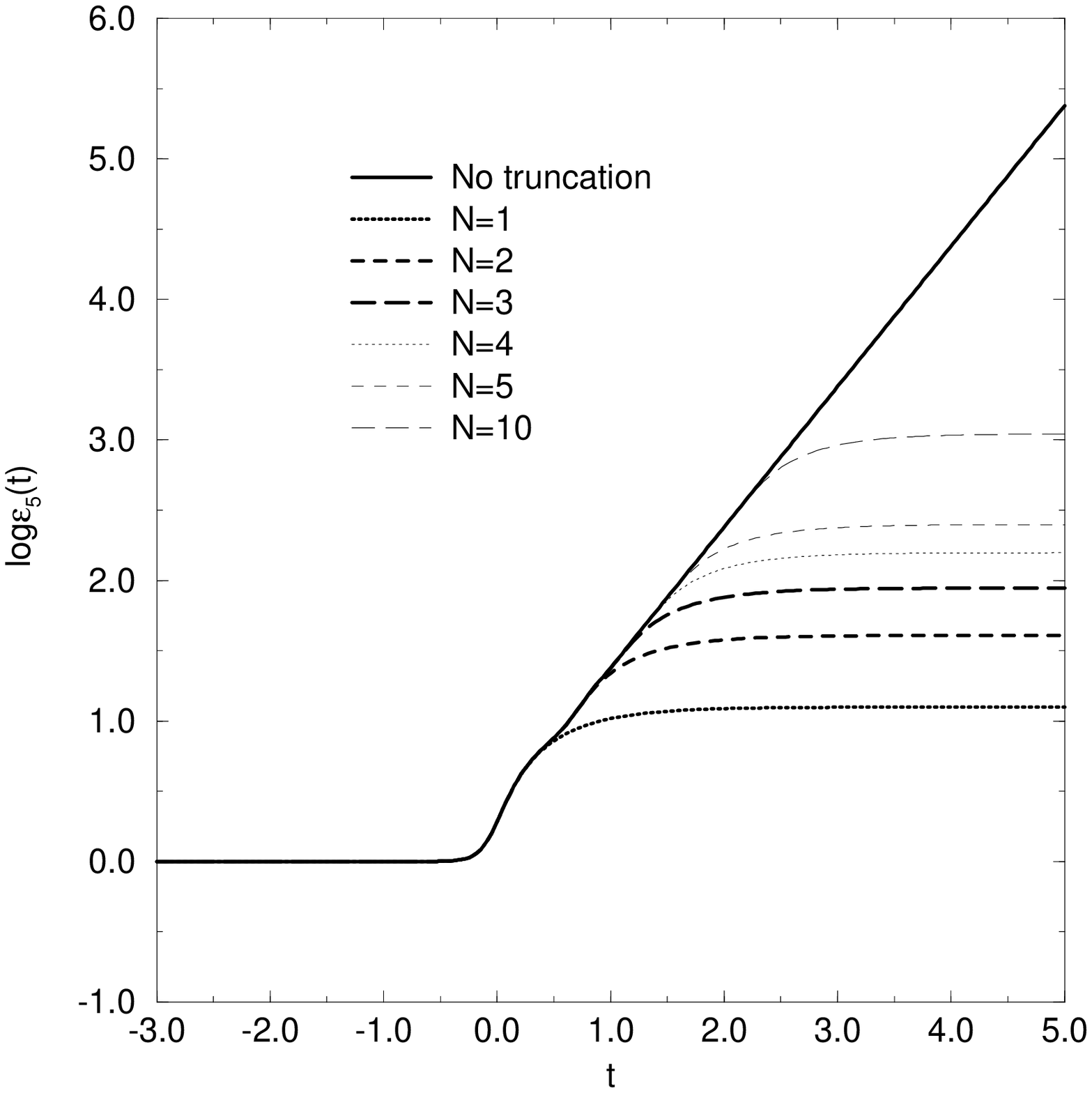}}
\label{fig:2}
\caption{The same as the fig.~1 with $k=5$.}
\end{figure*}

The asymptotic behavior of the $\beta$-function can be read off from 
figs.~1 and 2, and we see that it agrees with the expectation presented in 
(\ref{beta3}). 
In the present case, the $\beta$-function obeys the power law
behavior starting around $(R\Lambda) \sim 1$ (i.e.  $t \sim 0$).
As we can see from figs.~1 and 2 the function $\epsilon_k(t)$ 
interpolates the two regions of energy, below and above
the compactification scale $1/R$.
In fact, we may regard 
\be
D_{\rm eff} (R\Lambda) =
4+\frac{d \ln \epsilon_k}{d\ln (R\Lambda)}
\ee
as the effective dimension
\footnote{
The notion of the effective dimension here has been introduced 
in ref.~\cite{connor} 
to absorb the environmental effects (e.g. temperature)
occurring in dimensional
crossover phenomena into a redefinition of the spatial dimension.},
because the redefined coupling by 
\be
h_k &=& ~\epsilon_k(R\Lambda) \lambda~
\label{deff1}
\ee
satisfies the RG equation
\footnote{Indeed the expansion parameter in high energy region
is not the four dimensional coupling $\lambda$, but this
re-defined coupling $h_k$. However this coupling rapidly grows so that
the perturbative treatment may not be justified. 
In gauge theories the re-defined non-abelian 
gauge coupling will also grow into non-perturbative region
in the high energy limit. However,
$N=2$ supersymmetry for the massive Kaluza-Klein modes 
might overcome this difficulty 
\cite{kakushadze2}.
};
\be
\Lambda \frac{dh_k}{d \Lambda} &=& 
(D_{\rm eff}-4) h_k+ \frac{3}{16\pi^2}~h_k^2~.
\ee
One indeed finds that
\be
D_{\rm eff} (R\Lambda) ~\to~ 
\left\{ \begin{array}{c}
4\\5  \end{array}\right.~~\mbox{  as}~~  R\Lambda ~\to~
\left\{ \begin{array}{c}
0\\ \infty \end{array}\right.~.
\ee
This is demonstrated in figs.~3 and 4 with $k=1$ and $5$,
respectively (where the sum in eq.~(\ref{epsilon1})
is truncated at $N$).
\begin{figure*}[p]
\epsfysize=0.4\textheight   
\centerline{\epsffile{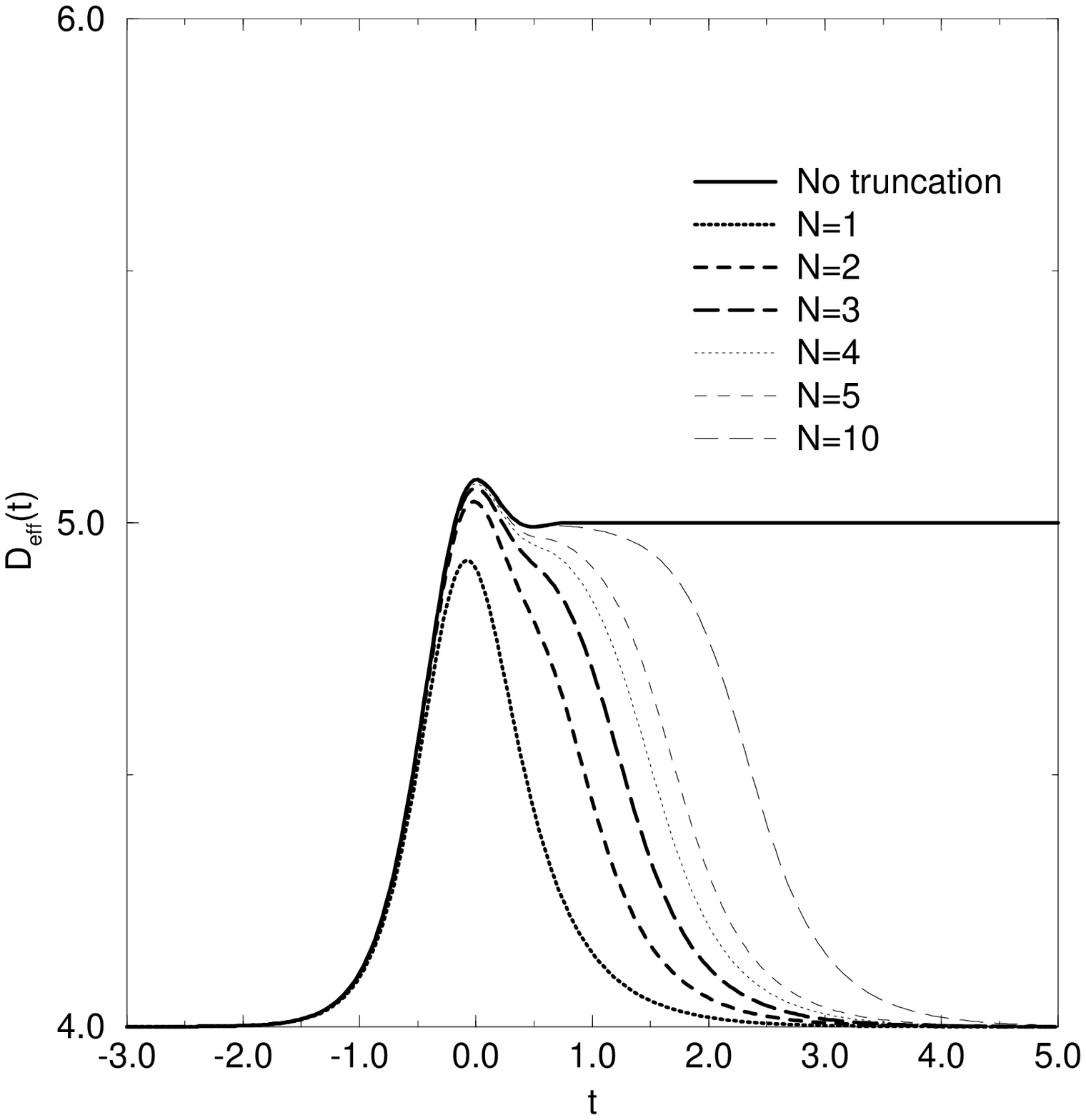}}
\caption{
The effective dimension $D_{\rm eff}(R\Lambda)$ for $k=1$ where 
$t=\ln (R\Lambda)$.}
\label{fig: 3}
\vspace{5mm}
\epsfysize=0.4\textheight   
\centerline{\epsffile{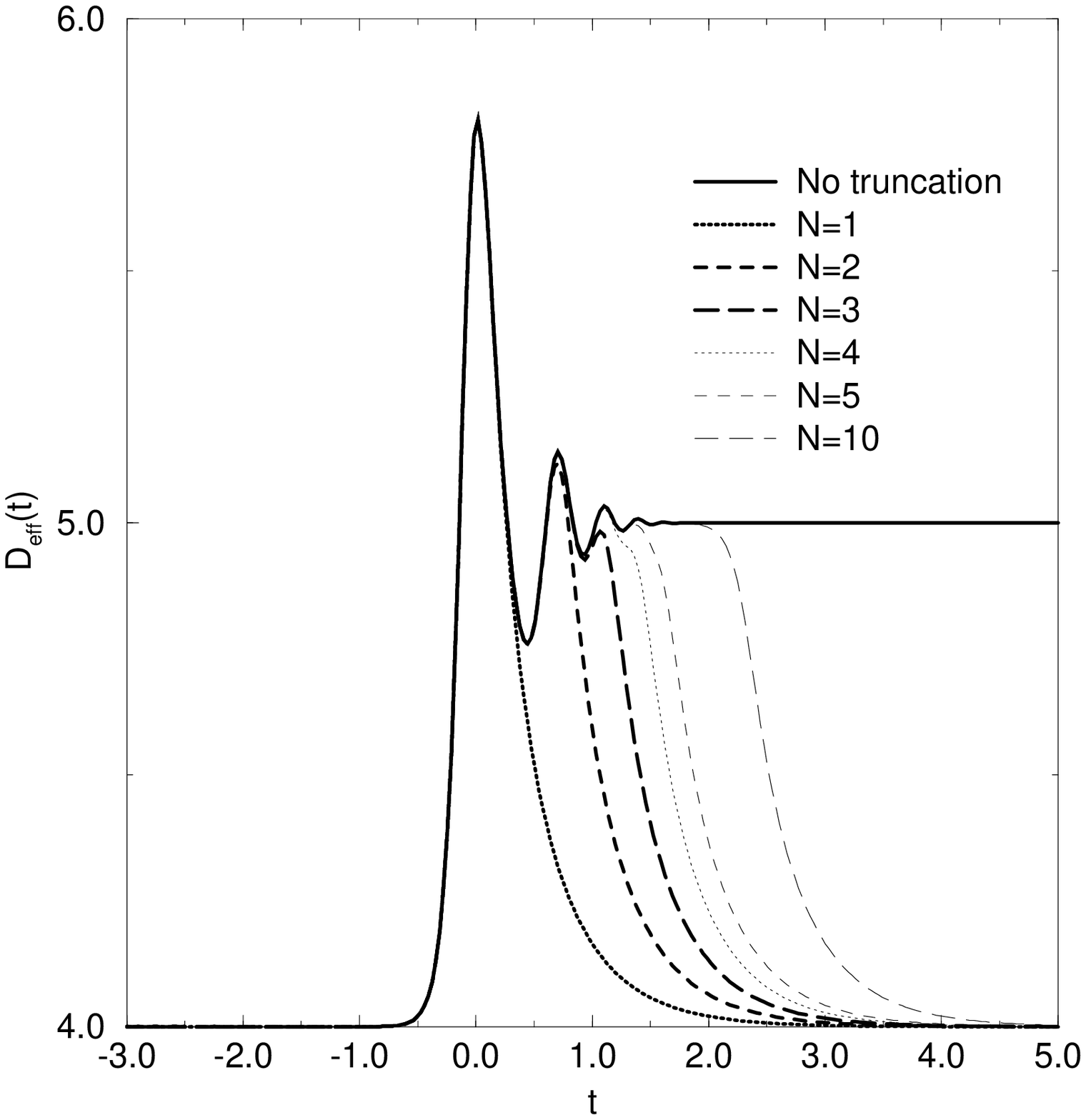}}
\caption{The same as fig.~3 with $k=5$.}
\label{fig: 4}
\end{figure*}
At this point we emphasize that the function $\epsilon_k$
contains the full information of the threshold effects of the 
Kaluza-Klein excitations. 
That is, the solution of the evolution
eq.~(\ref{rge-l3}) contains all the threshold effects.

Another important point is that the behavior of the effective 
dimension $D_{\rm eff}$
(which contains the infinite sum over the  Kaluza-Klein excitations, 
see  eq.~(\ref{epsilon1}))
near the compactification scale $1/R$ 
(in which the power law behavior of the coupling is not yet manifest) 
can  be well approximated by the sum over only the first few 
Kaluza-Klein excitations.
We see from figs.~2,~3 and 4 that only the first few Kaluza-Klein
excitations give important contributions to the threshold effects.
That is, it is not necessary to carry out the infinite sum in order to
find the form of the correction coming from the Kaluza-Klein threshold 
effects.
The coupling has the power law behavior already after passing few 
Kaluza-Klein excitations, and consequently, the $\beta$-function has 
reached  its asymptotic form given in eq.~(\ref{beta3}). 
That is, the Kaluza-Klein excitations should be indeed present for the 
coupling to obey the power law, but the asymptotic behavior, 
i.e. $b_2$ given in (\ref{beta3}), can be fixed through the first few 
excitations.

As a last task of this subsection we explicitly calculate the 
asymptotic form of the $\beta$-function for two different $k$'s,
where $k$ specifies the cutoff function $C$ given in
eq.~(\ref{cutoff4}) in the ERG scheme. 
We find
\be
(16\pi^2) b_2 = \left\{ 
\begin{array}{l}
5.5536\cdots \\
4.3832\cdots
\end{array}\right. 
~~\mbox{for} ~~
k=\left\{ 
\begin{array}{l}
1\\5  
\end{array}\right.~,
\label{deff2}
\ee
where the asymptotic coefficient $b_2$ is
defined in eq.~(\ref{beta3}).

\subsection{Momentum subtraction scheme}
Here we analyze the momentum subtraction (MOM) scheme 
\footnote{The MOM scheme has been applied
in early days of GUTs to take into account the
threshold effects of the super heavy particles
of GUTs into the evolution of the gauge
couplings \cite{d-ross}.}.
To this end, we  compute in the MOM scheme
the one-loop correction $\Pi^{(4)}_{\rm MOM}$ to the four point vertex
function in  the four dimensional theory defined
by the action in eq.~(\ref{action4}).
We find that for the massless case
\be
& &\Pi^{(4)}_{\rm MOM}(q^2,\Lambda^2=Q^2,m_n^2) \nn\\
&=&
\frac{3}{2}~\lambda^2~\int \frac{d^4 p}{(2\pi)^4}~
\sum_{n}
~\left[~\frac{1}{(p^2+m_n^2)((p-q)^2+m_n^2)}-
(q \to Q)~\right]~,
\label{pimom}
\ee
from which we obtain the one-loop $\beta$-function in the MOM scheme;
\be
\beta_{\rm MOM}
= \Lambda \frac{\partial 
\Pi_{\rm MOM}}{\partial \Lambda}
= \frac{3}{16\pi^2} \epsilon_{\rm MOM}(R\Lambda)~\lambda^2,
\ee
where
\be
\epsilon_{\rm MOM} (R\Lambda) = 
1+\sum_{n \neq 0}
\int_0^1 d x 
\frac{x (1-x) (R\Lambda/n)^2}{1+x (1-x) (R\Lambda/n)^2}~.
\ee
As we did in the case of the ERG scheme (see eq.~(\ref{deff1})),
we introduce the effective dimension in the MOM scheme:
\be
D_{\rm eff}=4+\frac{d \ln\epsilon_{\rm MOM}  }{d \ln (R\Lambda)}~.
\ee
In figs.~5 and 6 we plot $\ln\epsilon_{\rm MOM}$ and the 
$D_{\rm eff}$ in the MOM scheme
as a function of $t=\ln (R\Lambda) $.
\begin{figure*}[p]
\epsfysize=0.4\textheight   
\centerline{\epsffile{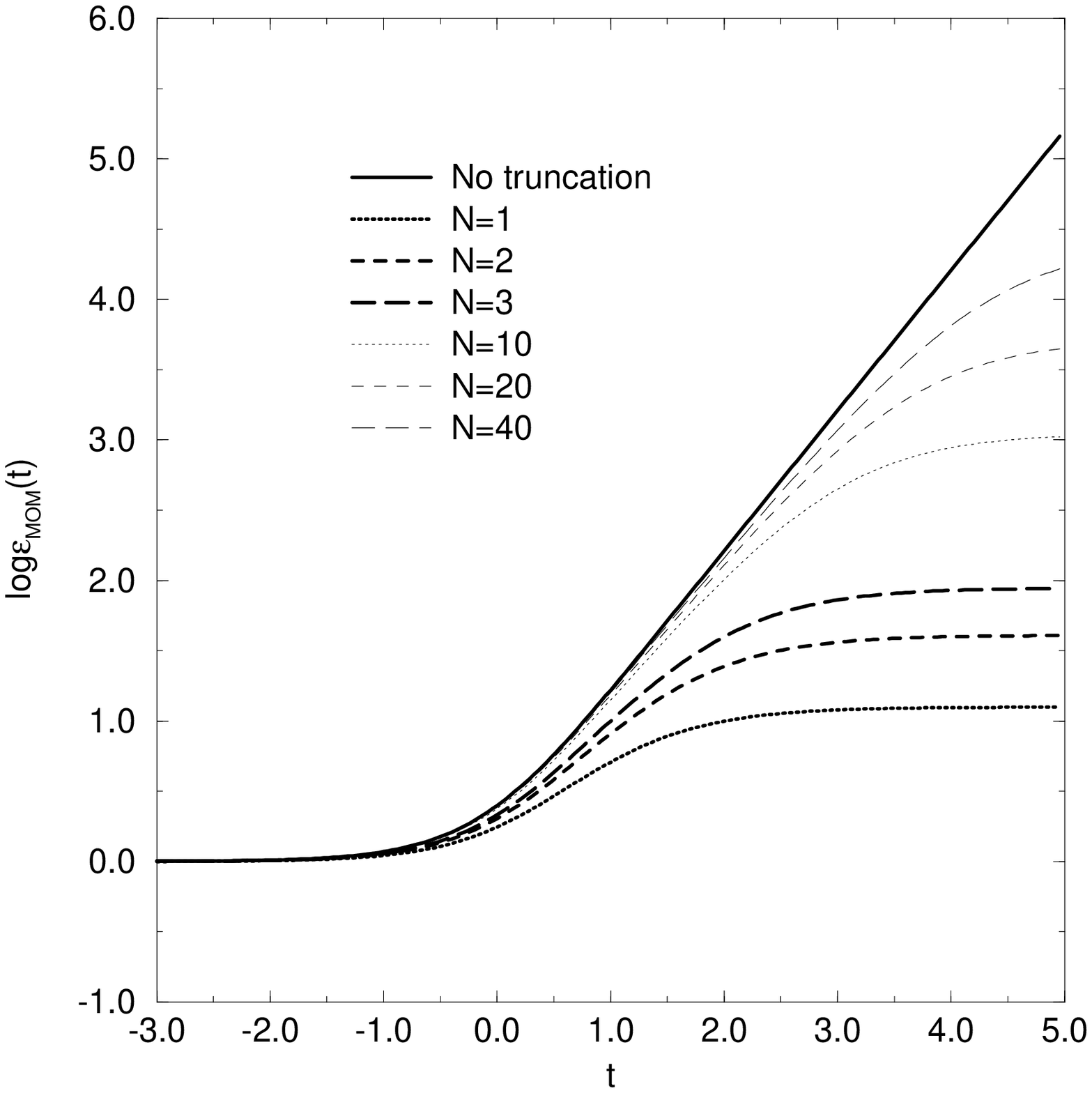}}
\caption{$\ln\epsilon_{\rm MOM}$ as a function of $t=\ln(R\Lambda)$.}
\label{fig: 5}
\vspace{5mm}
\epsfysize=0.4\textheight   
\centerline{\epsffile{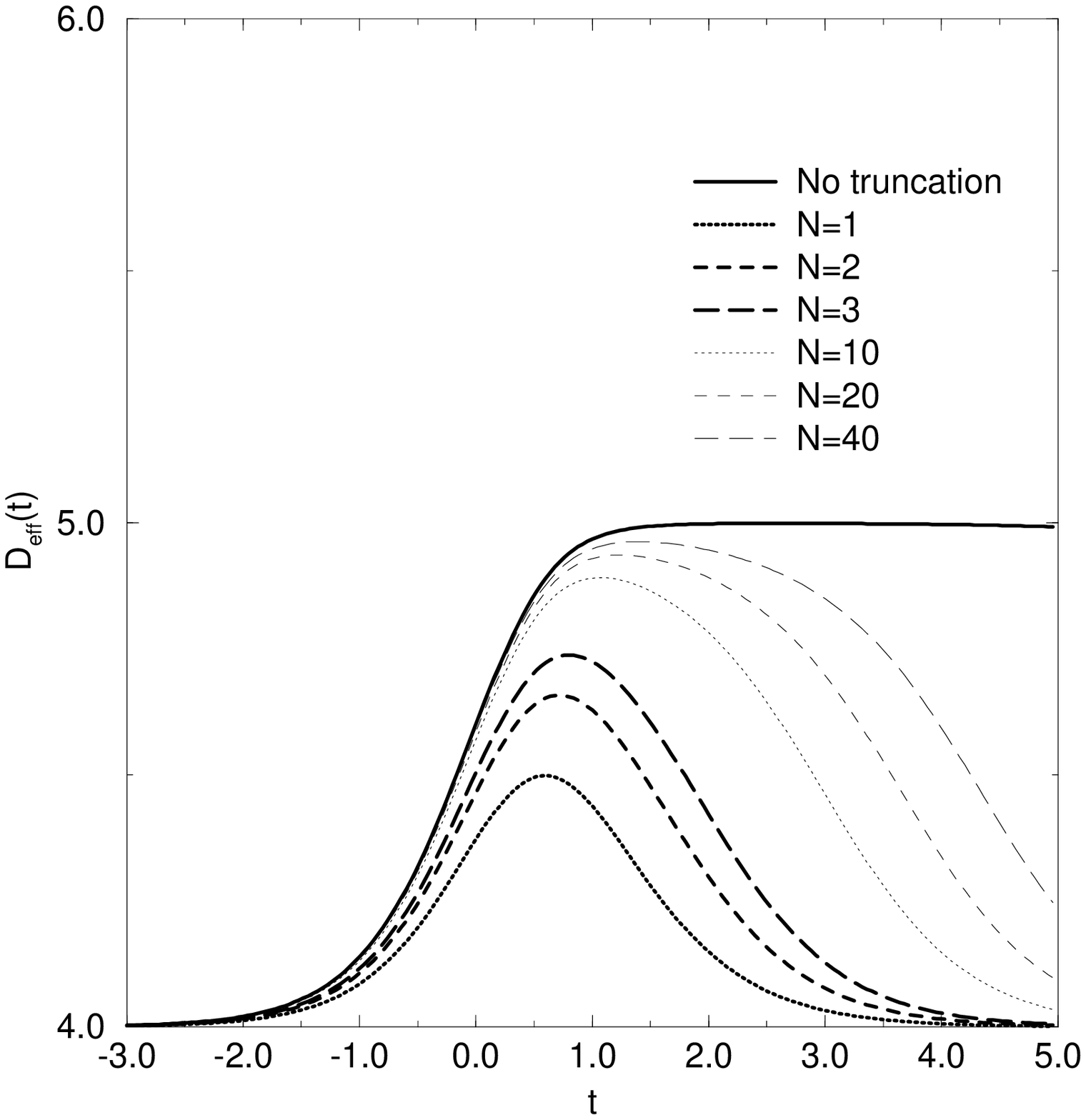}}
\caption{The effective dimension $D_{\rm eff}$ as a function of 
$t=\ln (R\Lambda)$.}
\label{fig: 6}
\end{figure*}
We see from figs.~5 and 6 that the power law behavior of
the $\beta$-function in the MOM scheme can take place for
$ R\Lambda \sim 2$ (i.e. $t \sim 0.7$).
The asymptotic form in the MOM scheme is found to be:
\be
 16 \pi^2 b_2 =3.0701\cdots~.
\label{deff3}
\ee

However, in contrast to the non-perturbative RG approach,
the transition region (i.e. $t \sim 0$) cannot
be approximated by the sum over only the first few excitations;
we need to sum over $\sim 40$ excitations, as is
demonstrated in fig.~6. Therefore, we can
conclude that  the MOM scheme is not an
economical one from this point of view.

\subsection{Proper time scheme}
Next we consider the proper time (PT) scheme.
This is the approximation used in refs.~\cite{veneziano,dienes1}.
So following them, we  compute as in the MOM scheme
the one-loop correction $\Pi^{(4)}$ to the four point vertex function
with zero external momenta in the PT scheme, and  find 
\be
\Pi^{(4)}_{{\rm PT},(r)} 
&=&\frac{3}{2}~\lambda^2~\int^{\prime} \frac{d^4 p}{(2\pi)^4}~
\sum_{n}\frac{1}{(p^2+m_n^2)^2}\
\label{pipt} \\
&=&\frac{\lambda^2}{8\pi^2}\frac{3}{2}
\int^{\mu^{-2}}_{r\Lambda^{-2}}\frac{dt}{t}~\left(\frac{1}{2}\right)~
~\vartheta_3\left(\frac{i t}{\pi R^2}\right)~, 
\ee
where $\vartheta_3$ is one of the Jacobi theta functions
\be
\vartheta_3 (\tau) &=&
\sum_{n=-\infty}^{\infty}\exp (i \pi n^2\tau)~,
\label{theta3}
\ee
and $\prime$ means that the $p$ integral is cutoff.
Note that an ultraviolet and infrared cutoff (by scale $\mu$) are 
introduced to make the $t$ integral finite, though the beta function 
is independent of the infrared cutoff. 
We emphasize that the factor $r$ cannot be fixed within the framework 
of the four dimensional theory, and so it is an ambiguity that belongs 
to the regularization scheme dependence uncertainties. 
In ref.~\cite{dienes1} a specific value was given;
\be
r &=& \pi (X_{\delta})^{-2/\delta}~,~
X_{\delta}=\frac{\pi^{\delta/2}}{\Gamma(1+\delta/2)}~,
\label{xdelta}
\ee
which has been obtained
 by comparing the asymptotic formula of 
$\vartheta_3$ (given in eq.~(\ref{theta3}))
and the total number of the Kaluza-Klein states with masses squared 
smaller than $\Lambda^2$. In the case at hand ($\delta=1$)
we have $r=\pi/4$.
But the difference between $r=\pi/4$ and another value
can be completely transformed away by means of the lowest non-trivial 
order transformation given in eq.~(\ref{transform1}), as we have 
explicitly verified
\footnote{The infinite sum appearing in eq.~(\ref{pipt}) is finite
in this case ($D=5$) and it can indeed be carried out analytically.
(See for instance, ref.~\cite{antoniadis6}.
Similar sums appear in field theories at finite temperature.)
Then one explicitly verifies that the results obtained 
by the two different calculations differ from each other, even in the 
asymptotic regime. But this difference is only apparent.}.

The $\beta$-function in the PT scheme is then
\be
\beta_{{\rm PT},(r)}=
\Lambda \frac{\partial \Pi_{{\rm PT},(r)}}{\partial \Lambda}
= \frac{3}{16\pi^2} \epsilon_{{\rm PT},(r)}(R\Lambda)~\lambda^2~,
\label{beta-pt}
\ee
where
\be
\epsilon_{{\rm PT},(r)}(R\Lambda)
=\vartheta_3 \left(\frac{i r }{\pi (R\Lambda)^2}\right) =
1+\sum_{n \neq 0}\exp\left(-\frac{ r n^2}{ (R\Lambda)^2}\right)~.
\label{epsilon-pt}
\ee
In fig.~7 and 8 we plot, respectively, $\ln\epsilon_{\rm PT,r}$ and the 
effective dimension
\be
D_{\rm eff}(R\Lambda) = 
4+\frac{d \ln\epsilon_{{\rm PT},(r)}(R\Lambda)  }
{d \ln (R\Lambda)}
\ee
as a function of $t=\ln (R\Lambda)$. for $r=\pi/4$.

We see from figs.7 and 8 that the power law behavior of
the coupling in the PT scheme can become manifest even before
$R\Lambda \sim 1$ (i.e. $t\sim 0 $).
Moreover, the threshold region (i.e. $R\Lambda \sim 1$) can
be very well approximated by the sum over the first few excitations,
which is shown in fig.~8.
\begin{figure*}[p]
\epsfysize=0.4\textheight   
\centerline{\epsffile{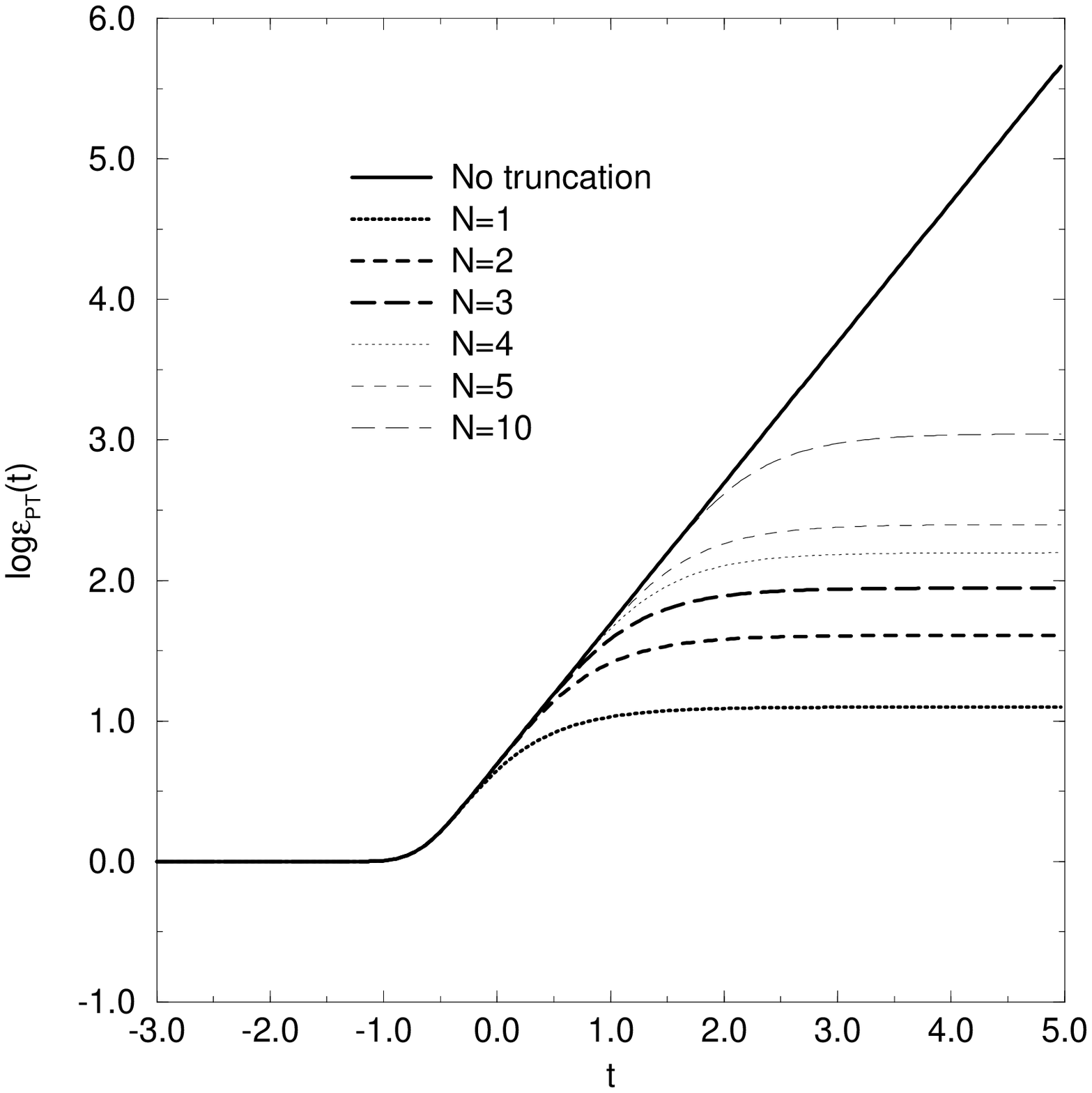}}
\caption{$\ln\epsilon_{\rm PT}$ with $r=\pi/4$ as a function of 
$t=\ln(R\Lambda)$ in the PT scheme.}
\label{fig: 7}
\vspace{5mm}
\epsfysize=0.4\textheight   
\centerline{\epsffile{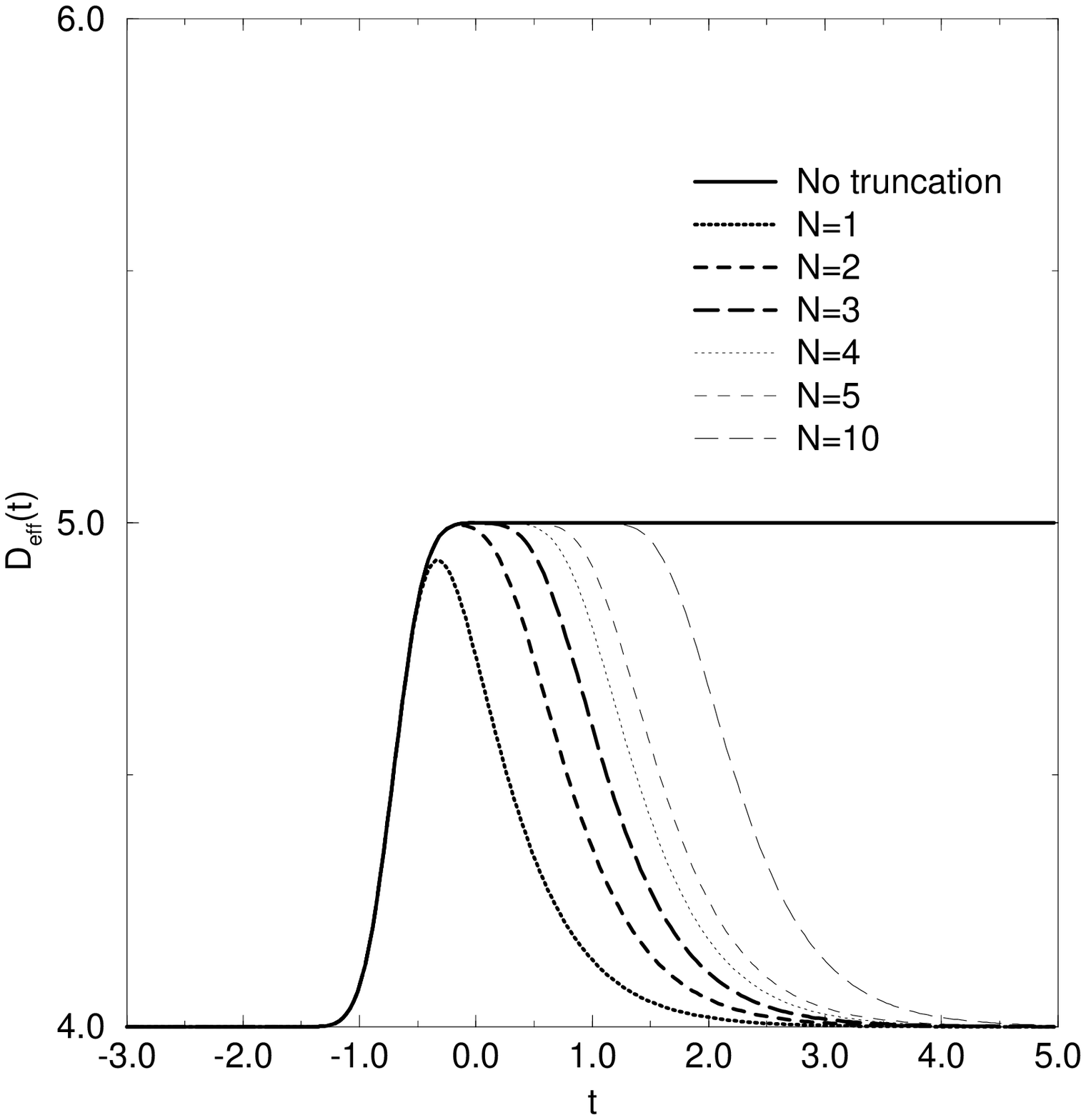}}
\caption{The effective dimension
$D_{\rm eff}$ as a function of $t=\ln (R\Lambda)$
for the PT scheme.}
\label{fig: 8}
\end{figure*}
The asymptotic form in the PT scheme (with $r=\pi/4$) turns out be
\be
16 \pi^2 b_2 = 6 
\label{deff4}
\ee
where $b_2$ is defined in eq.~(\ref{beta3}).

\subsection{Two dimensional torus compactification}
The investigation performed so far may be extended to the case
with  compactified dimensions more than one. 
(The ERG scheme and the proper time scheme are applicable for any 
dimensions. However, as for the momentum subtraction scheme defined by 
eq.~(\ref{pimom}), the infinite sum of the Kaluza-Klein modes diverges
for more than one compactified dimensions.)
Here we will examine the behavior of the effective dimensions in
the $D=6$ case of the torus compactification with two 
hierarchically different
radii $R_1$ and $R_2$ $(R_1 \ll R_2)$.
The asymptotic behavior of the RG equation at high energy scale
$\Lambda$ will be given by
\be
\Lambda\frac{\partial \lambda}{\partial \Lambda}
 \sim b_2 \Lambda^{-2} \lambda^2 + \cdots,
\ee   
and the effective dimensionality will be $6$ in the present case.
However, it is expected that the effective dimension is $5$
 at the 
intermediate scale of $R_1 \ll \Lambda \ll R_2$, 
and that successive dimensional crossover,
$D_{\mbox{eff}} = 6 \to 5 \to 4$,
occurs as  the Kaluza-Klein  modes decouple successively.

Extension of the formulae for the $\beta$-function coefficients 
given by eq.~(\ref{epsilon1}) in the ERG scheme and by 
eq.~(\ref{epsilon-pt}) in the PT scheme to the $D=6$ case 
is straightforward;
in the ERG scheme
\be
\epsilon_k(R_1\Lambda, R_2\Lambda)
&=&2(k+1)\sum_{n_1}\sum_{n_2} \int_0^{\infty}
ds s \frac{s_{n_1, n_2}^{2k}}{(1 + s_{n_1,n_2}^{k+1})^3},\nn \\
s_{n_1,n_2}&=&s + \left(\frac{n_1}{R_1 \Lambda}\right)^2
+\left(\frac{n_2}{R_2 \Lambda}\right)^2,
\ee
and in the PT scheme 
\be
\epsilon_{\mbox{PT},(r)}(R_1\Lambda, R_2\Lambda)
&=&\sum_{n_1}\sum_{n_2} \exp\left( -r\left[
\left(\frac{n_1}{R_1 \Lambda}\right)^2
+\left(\frac{n_2}{R_2 \Lambda}\right)^2
\right]\right)~ \nonumber \\
&=&\vartheta_3\left(\frac{ir}{\pi(R_1 \Lambda)^2} \right)~
\vartheta_3\left(\frac{ir}{\pi(R_2 \Lambda)^2} \right)~.
\ee
The prescription of ref.~\cite{dienes1} gives $r=1$ in this case.  
The effective dimension may be also defined from these functions
as  
\be
D_{\mbox{eff}}(R_1\Lambda)
=4 + \frac{d \ln\epsilon_i(R_1\Lambda, R_2/R_1)}{d\ln(R_1\Lambda)}.
\ee
In fig.~9 the effective dimension $D_{\mbox{eff}}$ as a function of 
$t=\ln(R_1\Lambda)$ in the torus compactification is shown, where
the radius ratio of the torus is set equal to $R_2/R_1=10$. 
Results are obtained for $k=5$ in the ERG scheme 
and $r=1$ in the PT scheme.
\begin{figure}[htb]
\epsfysize=0.4\textheight   
\centerline{\epsffile{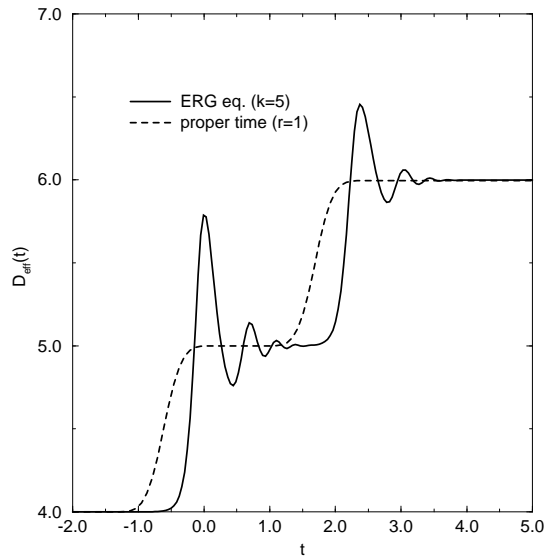}}
\caption{
The effective dimension $D_{\mbox{eff}}$ as a function of 
$t=\ln(R_1\Lambda)$ in the case that the compactified dimension is 
a torus with radius ratio of $R_2/R_1=10$. 
Results in the ERG scheme $(k=5)$ and in the PT scheme are shown.
}
\label{fig:9}
\end{figure}
The successive dimensional crossover  is presented in fig.~9. It is
noted that the PT scheme gives a smooth transition of the
effective dimensionality.

\section{Results and applications}

\subsection{Comparison}
We have explicitly seen in the previous section that all the
regularization schemes we have considered belong to the equivalent
class of the schemes in which the asymptotic limit of the coupling 
obeys the same power law.
This is the basis which enables us to make a meaningful comparison of
the results obtained in the different schemes.
Since we stay within the one-loop approximation, we may assume that 
the coupling constant $\lambda$ in all the schemes has the same 
value, say $\lambda_0$, at $\Lambda_0$ which is very much smaller 
than $1/R$.
Furthermore, for $\Lambda$ much below $1/R$, all the schemes have the 
same one-loop $\beta$-function, and the difference among the different 
schemes must become larger and larger when closing to the transition 
region.
After the transition region the difference of the schemes can be well
controlled, and the result obtained in one scheme can be
transformed into the equivalent one 
in such a way that they can be meaningfully compared.

To proceed, we define a scheme which has no threshold correction
(the one-$\theta$ function approximation):
\be
\beta_0 &=& \frac{1}{16\pi^2}\lambda^2~[~
3 \theta(1-R\Lambda) +6 \theta(R\Lambda-1)
(R\Lambda)~]~,
\label{beta-0}
\ee
where we have used as the coefficient for the second term 
the one suggested in ref.~\cite{dienes1}.
That is, the scheme is specified by
\be
(16\pi^2) b_2 &= &6 ~.
\label{zero}
\ee
We would like to transform the result obtained, say in the $i$-scheme 
specified by $16 \pi^2 b_{2 i}$,
to a reference  scheme for which we choose 
the PT scheme which is specified by $16 \pi^2 b_2=6$
as in the case of the one-$\theta$ function scheme.
According to eq.~(\ref{transform1}), the transformation has the form
\be
\lambda_i ~\to~
\lambda_i +c_{2 i} \lambda^2 \Lambda~.
\label{transform3}
\ee
The subscript $i$  should indicate that 
the actual values of the transformed $\lambda_i$ differ
from one scheme to another, although all the schemes have been brought 
into the same scheme through the transformation (\ref{transform3}) 
upon  using eqs.~(\ref{beta5}).
This difference will
be the net difference of the regularization schemes.

We use the initial condition
\be
\lambda = \lambda_0 =0.6  ~~\mbox{at}~
~t_0=\ln(R\Lambda_0) =-3~,
\label{initial1}
\ee
and we solve the evolution equation for $\lambda$ in the different 
schemes.
Fig.~10 shows the $\beta$-functions of the different schemes
after they have been brought to the reference scheme, the PT scheme.
\begin{figure}[htb]
\epsfysize=0.4\textheight   
\centerline{\epsffile{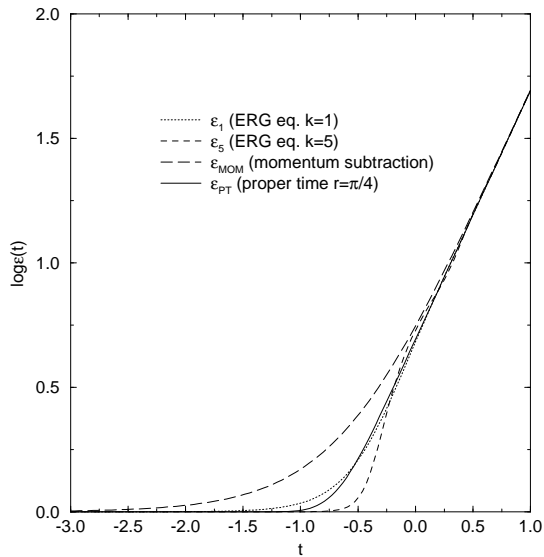}}
\caption{The $\ln\epsilon$ in the different schemes after the 
appropriately chosen finite transformation.}
\label{fig: 10}
\end{figure}
From fig.~10 we see that the ERG scheme and the PT scheme are
very similar and that the difference in the threshold effect
in the different schemes is visible only for $t \lsim 1$.
Fig.~11 shows the evolution of the coupling $\lambda_i$ as a function of
$t=\ln (R\Lambda)$ varying from $-2$ to $3$ in the different schemes 
including the one-$\theta$ function scheme defined by
eq.~(\ref{beta-0})
\footnote{
In figs.~11 and 12 we have included for comparison also 
the successive-$\theta$ function approximation scheme, 
which represents a completely
renormalizable four dimensional theory with successive discrete
thresholds of massive Kaluza-Klein states. This scheme actually
suggests $r$ given in eq.~(\ref{xdelta}) to be $\pi/4$.
}.
As we see can see from fig.~11 that the difference is very small.
To see the difference we present the evolution in the threshold 
region (i.e. $-1.0 \lsim t \lsim 1.0$) in fig.~12.
\begin{figure}[htp]
\epsfxsize=0.4\textheight 
\centerline{\epsffile{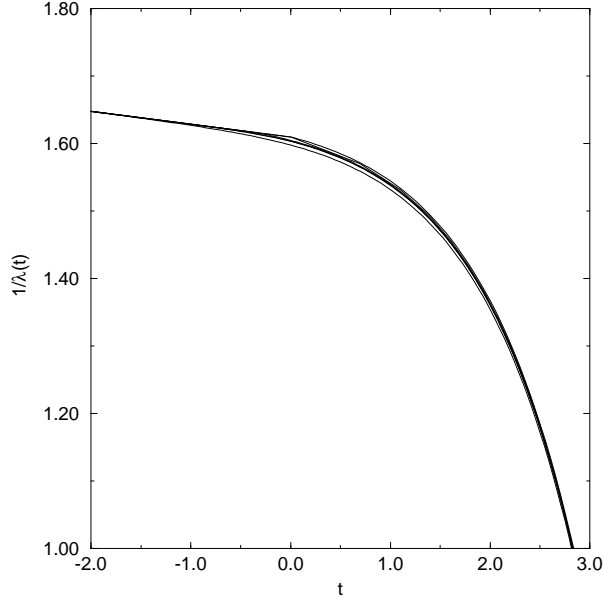}}
\caption{The evolution of $\lambda$ in the various schemes.}
\label{fig: 11}
\end{figure}
\begin{figure}[htp]
\epsfxsize=0.8\textheight 
\centerline{\epsffile{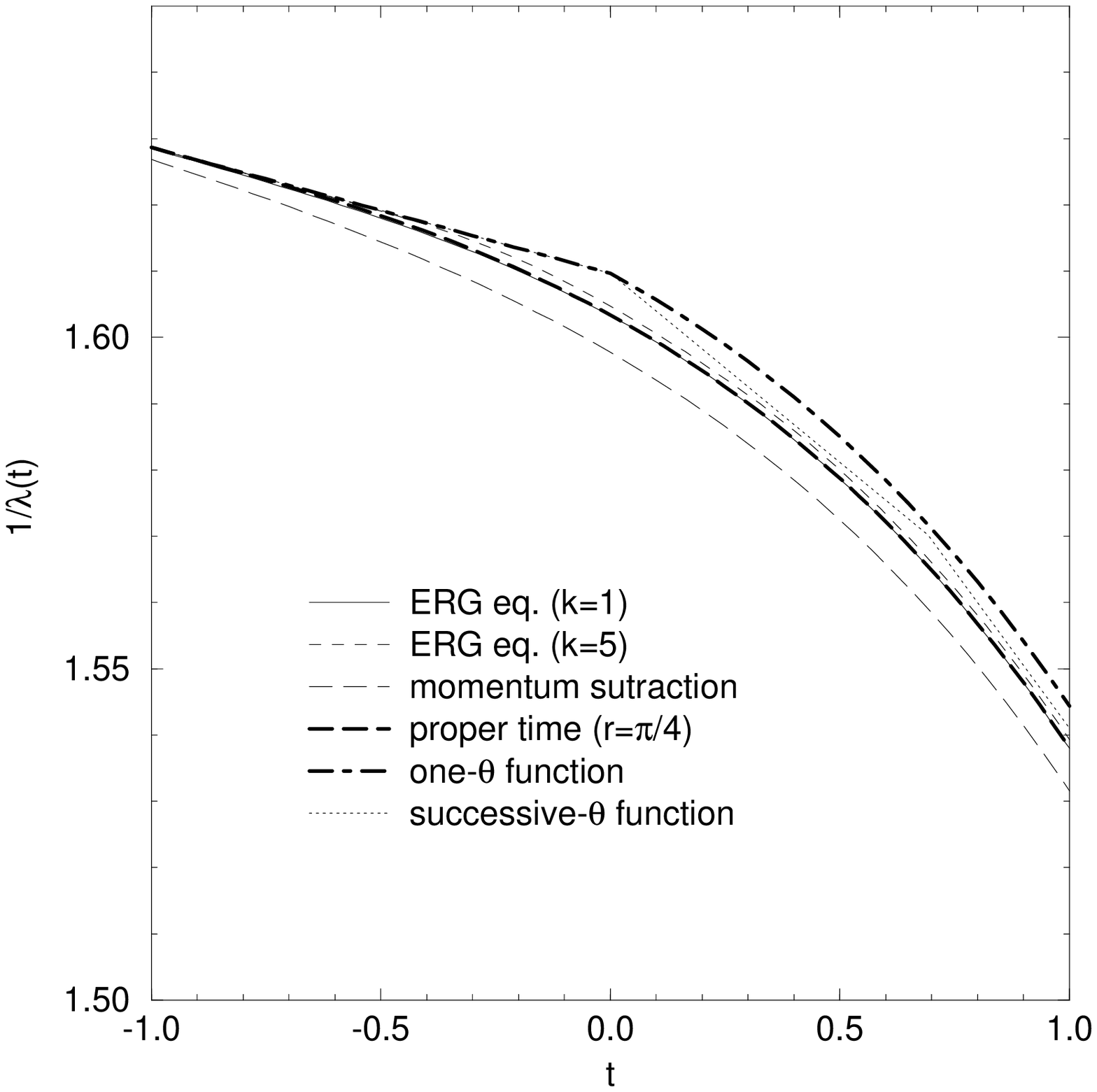}}
\caption{A magnified figure of fig.~11 near the threshold region.}
\label{fig: 12}
\end{figure}
So the maximal difference in this regime is $O(0.8 )(\lambda/4\pi)$.
Above $t=2$, which means above the $7$th Kaluza-Klein excitation,
the maximal difference including the zero scheme
is $O(0.08 )(\lambda/4\pi)$.

\subsection{Correction to gauge coupling unification}

Now we would like to apply our result in the previous sections
to the running of the gauge couplings. Here we will consider
only the PT  and  one-$\theta$ function approximation schemes,
because, as we have seen,
the other schemes yield similar results.
Suppose that the one-loop $\beta$-function of a gauge coupling
$g_{\rm PT}$ has the form
\be
16 \pi^2 \beta_{\rm PT} =
g^3_{\rm PT} \left\{
~b + \tilde{b} ~\left[~
\vartheta_3^{\delta}\left(\frac{i r}{\pi (R\Lambda)^2}\right) -1~
\right]~\right\}
\label{beta61}
\ee
in the PT scheme with the normalization used in ref.~\cite{dienes1}
(i.e. $ r= \pi (X_{\delta})^{-2/\delta}$
with $X_{\delta}=\pi^{\delta/2}/\Gamma(1+\delta/2)$),
where the function $ \vartheta_3(i r/\pi (R\Lambda)^2)$
is given in eq.~(\ref{epsilon-pt}).
The constant $b$ accounts for the massless modes
and $\tilde{b}$ for the massive Kaluza-Klein modes.
Since, as we can see in fig.~7,  
$\vartheta_3(i r/\pi (R\Lambda)^2)$
for $\ln(R \Lambda) \gsim T$  can be well approximated
$ (\pi/r)^{\delta/2}(R\Lambda)^{\delta}$,
we replace the r.h.side of eq.~(\ref{beta61}) by
\be
16\pi^2 \beta_{\rm PT}
&\simeq& g^3_{\rm PT}\left\{b-\tilde{b}+
\tilde{b}\left[
\vartheta_3^{\delta}\left(\frac{i r}{\pi (R\Lambda)^2}\right)
\theta(T-\ln (R\Lambda))
\right. \right. \nn\\
& & \left. \left.
+ \left(\frac{\pi}{r}\right)^{\delta/2}(R\Lambda)^{\delta} 
\theta(\ln (R\Lambda)-T)
\right]\right\}~,
\ee
where $T$ is some value $\gsim 0$ in the asymptotic regime.
This $\beta$-function yields 
 the evolution of $g_{\rm PT}$ 
\be
\frac{16\pi^2}{g_{\rm PT}^2(\Lambda)}-
\frac{16 \pi^2 }{g_{\rm PT}^2(\Lambda_0)}
\simeq  -b \ln \frac{\Lambda}{\Lambda_0}
+\tilde{b}\ln (R\Lambda)
-\tilde{b}\left\{\frac{(\pi/r)^{\delta/2}}{\delta}[
(R\Lambda)^{\delta}-1]+\Delta_{\delta}\right\},
\label{evol4}
\ee
for $\Lambda> (1/R) e^T > 1/R$, where
\be
\Delta_{\delta} = T+
\frac{(\pi/r)^{T\delta/2}}{\delta}(1-e^{\delta})
+\int_{-T_0}^{T}dt 
\left[~
\vartheta_3^{\delta}\left(\frac{i r e^{-2 t}}{\pi}\right)-1~
\right]~,
\label{deltadef}
\ee
where we can replace $T_0$ by $-\infty$ as
long as $T_0 \ll 0$.
Since $T$ is in the asymptotic regime, $\Delta_{\delta} $ 
does not depend on $T$.
The $\Delta_{\delta}$ is the Kaluza-Klein threshold effect and
we find for instance
\be
\Delta_{\delta}( r= \pi (X_{\delta})^{-2/\delta} ) &\simeq &
\left\{ \begin{array}{c}
0.33\\0.55\end{array}\right. ~~\mbox{for} ~~\delta=
\left\{ \begin{array}{c}
1\\2  \end{array}\right.~.
\label{delta21}
\ee
It depends however on $r$, e.g.
\be
\Delta_1 (r=1) &\simeq &0.22 ~,~\Delta_1 (r=2) \simeq  0.050
~,~\Delta_1 (r=\pi) \simeq 0.023 ~.
\ee
$\Delta_{\delta}$  decreases for 
increasing $r$ and reaches at its minimum 
for $r=\pi$. So the $\pi$-scheme (defined by 
$r=\pi$) might be the most economic scheme,
because one might neglect the threshold effect completely
in this scheme.
This can be seen by calculating the difference
\be
\Delta_{\delta}(r)-\Delta_{\delta}(r')
&=&\frac{1}{\delta}~[~(\frac{\pi}{r})^{\delta/2}
 -(\frac{\pi}{r'})^{\delta/2} ~]+
\frac{1}{2}\ln\frac{r}{r'}~.
\label{DD}
\ee
Differentiating the r.h.side with respect to $r$,
we find that $\Delta_{\delta}(r)$ becomes minimal
at $r=\pi$. This dependence of $r$ is unphysical because we
can absorb it into a redefinition of the coupling,
or equivalently into a redefinition of $\Lambda$, as we will see.

It is now straightforward to apply the above 
formula to investigate the threshold effect to  gauge
coupling unification. To be definite, we consider the 
Kaluza-Klein model proposed in ref.~\cite{dienes1},  
the MSSM with the  Kaluza-Klein towers only in the gauge 
and  Higgs supermultiplets.
The gauge coupling $\beta$-functions are given by \cite{dienes1}
\be
\left\{ \begin{array}{l}
b_1=33/5\\b_2=1\\b_3=-3\end{array}\right. ~~\mbox{and}~~ 
\left\{ \begin{array}{l}
\tilde{b}_1=3/5 \\ \tilde{b}_2=-3 \\
\tilde{b}_3=-6 \end{array}\right.~.
\label{bs}
\ee
As usual we use the $\alpha_1(M_Z)~(=g_1^2(M_Z)/4\pi)$ and 
$\alpha_2(M_Z)$ as the input, and predict $\alpha_3(M_Z)$
and $M_{\rm GUT}  $.
We find that threshold effect changes the unification scale
as
\be
\left(\frac{b_1-b_2}{\tilde{b}_1-\tilde{b}_2}-1\right) 
\ln\frac{M_{\rm GUT}^{(\theta)}}{M_{\rm GUT}^{({\rm PT})}}
+\frac{(\pi/r)^{\delta}}{\delta}~R^{\delta}~
\left[~
(M_{\rm GUT}^{(\theta)})^{\delta}-
(M_{\rm GUT}^{({\rm PT})})^{\delta}~
\right]-\Delta_{\delta} = 0~,
\label{mgut}
\ee
where $ M_{\rm GUT}^{(i)}$ is the unification scale
in the $i$ scheme. 
Since $\Delta_{\delta}$ (which is given in
eq.~(\ref{deltadef})) and the factor in front of $\ln$
in eq.~(\ref{mgut}) are positive,
 the unification scale in the PT scheme with $r=\pi/4$ is 
smaller than the one without the Kaluza-Klein threshold effect.
For the model at hand the difference is only
$\sim 1$ \%. 

The prediction of $\alpha_3(M_Z)$ does not changes practically;
there is a compensating mechanism. As we have observed, 
$M_{\rm GUT}^{(\theta)}$  is slightly larger that 
$M_{\rm GUT}^{({\rm PT})}$, which means that 
$\alpha_3(1/R)$ in the one-$\theta$ function approximation
scheme with $M_{\rm GUT}^{(\theta)}$ is  larger as 
compared with that of $ M_{\rm GUT}^{({\rm PT})} $.
Similarly, 
$\alpha_3(1/R)$ in the PT scheme with $ M_{\rm GUT}^{({\rm PT})}$
is larger than the corresponding in the one-$\theta$ function
approximation  scheme with the same unification scale.
These two effects are of the same magnitude.
But it is clear that this compensating mechanism depends on the model.
Since the threshold effect is
\be
\delta\left(\frac{1}{\alpha}\right)
& \simeq& ~ \Delta_{\delta} ~\frac{\tilde{b} }{4\pi}~,
\ee
the effect could become in principle of the order of
few percent in the PT scheme with
$r=\pi/4$, {\em i.e.} comparable to the present 
experimental error \cite{pdg}.

The $r$-dependence in the prediction of $\alpha_3(M_Z)$ 
should be absent.
Let us see this explicitly at one-loop order.
We start with the assumption that the unification condition
\be
\frac{8\pi^2}{g_{{\rm GUT},r}^2(M_{\rm GUT}^{(r)})} &=&
\frac{8 \pi^2 }{g_{r,i}^2(\Lambda_0)}
 -b_i \ln \frac{M_{\rm GUT}^{(r)}}{\Lambda_0}
+\tilde{b}_i\ln (RM_{\rm GUT}^{(r)})\nn\\
& &-\tilde{b}_i\left\{\frac{(\pi/r)^{\delta/2}}{\delta}[
(RM_{\rm GUT}^{(r)})^{\delta}-1]+\Delta_{\delta}(r)~\right\}~~
\mbox{for all} ~~i~
\label{gut4}
\ee
is satisfied in the $r$ scheme,
where $\Lambda_0 < 1/R$ is the energy scale at which we
match the unrenormalizable
theory with  a renormalizable effective theory , say in
the $\overline{\mbox{MS}}$ scheme.
We now would like to prove
that the unification condition in the $r'$ scheme as well
is satisfied  and that the 
$g_{3}^2(\Lambda_0)$ prediction is $r$-independent.
Comparing the asymptotic behavior of the couplings of two schemes,
 we first find that the unification scales are related by
\be
M_{\rm GUT}^{(r')} &=&M_{\rm GUT}^{(r)} (\frac{r'}{r})^{1/2}~.
\ee
Then using the identity (\ref{DD}), we can rewrite 
the r.h.side of eq.~(\ref{gut4}) as
\be
& &\frac{8 \pi^2 }{g_{r,i}^2(\Lambda_0)}-
\frac{1}{2}b_i\ln \frac{r}{r'}
 -b_i \ln \frac{M_{\rm GUT}^{(r')}}{\Lambda_0}
+\tilde{b}_i\ln (RM_{\rm GUT}^{(r')})\nn\\
& &-\tilde{b}_i\left\{\frac{(\pi/r')^{\delta/2}}{\delta}[
(RM_{\rm GUT}^{(r')})^{\delta}-1]+\Delta_{\delta}
(r')~\right\}
+     ~~
\mbox{for all} ~~i~,
\label{gut5}
\ee
where the sum of the first two terms is nothing but
\be
\frac{8 \pi^2 }{g_{r',i}^2(\Lambda_0)}
&=&\frac{8 \pi^2 }{g_{r,i}^2(\Lambda_0)}-
\frac{1}{2}b_i\ln \frac{r}{r'}~,
\label{rr}
\ee
which can be obtained by computing
the corresponding diagrams 
for $\Lambda_0 < 1/R$ in two different PT schemes.
This shows the $r$-dependence in one-loop order.
To match 
$g_{r,i}(\Lambda_0)$ with 
$g_{\overline{\mbox{MS}}}(\Lambda_0)$, we still have to
clarify the relation between the 
$\overline{\mbox{MS}}$ scheme and the PT scheme
in the MSSM. This is certainly outside of the scope of
this paper, and we would like to leave it for future
work 
\footnote{For the scalar theory, however, we can give such
relation:
\be
\frac{16\pi^2}{\lambda_{r}(\Lambda_0)} &=&
\frac{16 \pi^2 }{\lambda_{\overline{\mbox{MS}}}(\Lambda_0)}
+\frac{b}{2} ~[~2+\frac{1}{6} r
-\int_{0}^{1} dx   \int_{r}^{\infty}
\frac{dt }{t} \exp -t x (1-x)~]~,
\ee
where the quantity in the parenthesis 
is $1.269\dots~ (0.2098\dots) ~\mbox{for}~ r=\pi (\pi/4)$.}.

\subsection{Comparing with the string Kaluza-Klein
threshold effect}

String theories in general contain different kinds 
of towers of massive modes;
the original massive string modes, 
the massive winding modes and the massive Kaluza-Klein modes.
Their characteristic mass scale is,
respectively, $O(1/\sqrt{\alpha'})$, $O(1/R)$
and $(R/\sqrt{\alpha'})O(1/\sqrt{\alpha'})$, where
$\alpha'$ is the Regge slope.
 These massive states give rise to
quantum contributions to the gauge couplings 
\footnote{See refs. \cite{kakushadze2,ross2}
and references therein.}, and
in the case of the hierarchal mass relation, that is, if
$1/R \ll 1/\sqrt{\alpha'} $,  the Kaluza-Klein states
are lighter than the others,  so that one can expect
that the Kaluza-Klein modes will give dominant
contributions to the quantum 
effect  \cite{kakushadze2,ross2}. To be definite, in what follows
we consider the weakly coupled heterotic string theory. 
The quantum corrections in this theory have been calculated
\cite{kaplunovsky,stringcorr},
and their form is given by 
\be
\frac{8\pi^2}{g_{s}^{2}} &=&
\frac{8\pi^2}{g_{{\rm ST},i}^{2}}(M_s) =
\frac{8\pi^2}{g_{{\rm ST},i}^{2}}(Q)
-b_i \ln \frac{M_s}{Q} - D_i~,
\ee
where $D_i$ stands for the threshold effect of the massive states,
and the string scale $M_s$ is related 
to $\alpha'$ through \cite{kaplunovsky}
\be
M_s &=& \frac{\zeta}{\sqrt{\alpha'}}~,~
\zeta=2\exp[(1-\gamma_E)/2] 3^{-3/4}(2\pi)^{-1/2}
\simeq 0.432~.
\label{ms}
\ee
We assume that the six dimensions are compactified on a
six dimensional orbifold, so that 
 the contribution to $D_i$ comes only from 
the massive supermultiplets forming a  $N=2$ but not $N=4$
 supermultiplet. In the hierarchal mass relation,
$1/R \ll 1/\sqrt{\alpha'} $, the contribution to $D_i$ for 
$Q \ll M_s$ is dominated 
by the Kaluza-Klein modes, and it can be written as \cite{ross2}
\be
D_i = \frac{1}{2}\tilde{b}_i \int_{-1/2}^{1/2}d \tau_1~
\int_{\sqrt{1-\tau_1^2}}^{\infty} \frac{d\tau_2}{\tau_2}~ 
\left\{ ~\sum_{m_1,m_2 \in {\bf Z}}\exp
\left[~
-\frac{\pi\tau_2\alpha'}{R^2}(m_1^2+m_2^2)~
\right]-1\right\}~,
\ee
where we have assumed that the radii
associated with the two-dimensional torus 
embedded  in the six-dimensional torus are both $R$, and
$\tau=\tau_1+i \tau_2$ is the modulus of the word sheet torus
corresponding to the one-loop word sheet topology.
Making the change of the variable, $\tau_2 \to t=\alpha' \tau_2$,
we obtain
\be
D_i &=&  \frac{1}{2}\tilde{b}_i \int_{-1/2}^{1/2}d \tau_1~
\int_{\alpha'\sqrt{1-\tau_1^2}}^{\infty}\frac{dt}{t}~ 
\left\{ ~\sum_{
m_1,m_2 \in {\bf Z}}\exp
\left[~-\frac{\pi t}{R^2}(m_1^2+m_2^2)~\right]-1\right\}\nn \\
&=& \frac{1}{2}
 \tilde{b}_i \int_{-1/2}^{1/2}d \tau_1~
\int_{\alpha'\sqrt{1-\tau_1^2}}^{\infty}\frac{dt}{t}~ 
\left\{
\vartheta_3^2\left(\frac{i t }{R^2}\right) -1\right\}~.
\label{di}
\ee
Then we  interpret eq.~(\ref{di}) as a result of the ``running''
of $g_{{\rm ST},i}$ from $1/R$ to $M_s=\zeta /\sqrt{\alpha'}$.
The corresponding $\beta$-function may be obtained from
\be
\beta_{{\rm ST},i}&=&
\Lambda \frac{d}{d \Lambda} g_{{\rm ST},i}(M_s=\Lambda)\nn\\
&=& \frac{g_{{\rm ST},i}^{3}}{16\pi^2}~
\left\{ b_i+
\tilde{b}_i \int_{-1/2}^{1/2}d \tau_1~
\left[~
~\vartheta_3^2 
\left(\frac{i \sqrt{1-\tau_1^2}\zeta^2 }{(\Lambda R)^2}\right) -1~
\right]~\right\},
\ee
where $\zeta$ is given in eq.~(\ref{ms}).
Comparing this with eq.~(\ref{beta61}) we see that the $\tau_1$ 
integral averages the 
field theory results in different regularizations 
with a definite weight.
Therefore, the string scheme corresponds to an average scheme,
and by investigating
the $\beta$-function in the asymptotic regime we can 
find the effective regularization scheme of, as we shall do below.
We find that for $R\Lambda \gg 1$
\be
\beta_{{\rm ST}, i} &\simeq & g_{{\rm ST}, i}^3 \{b_i+
\tilde{b}_i~G_{2}~(R\Lambda)^{2}~\},
\ee
where 
\be
~G_{\delta} &=& \frac{1}{\zeta^{\delta}}\int_{-1/2}^{1/2}d
\tau_1~(1-\tau_1^2)^{\delta/4}~=
 \frac{\pi}{3\zeta^2} \simeq 5.60~\mbox{for}~~\delta=2~.
\ee
According to the discussion in section 2 (see eq.~(\ref{beta4})),
the couplings in the PT scheme in
the normalization of ref.~\cite{dienes1}
($r=1$ for $\delta=2$) and in the string theory are 
related by
\be
\frac{8\pi^2}{g_{{\rm PT},}^2} &=& 
\frac{8\pi^2}{g_{{\rm ST},}^2} -
\frac{\pi}{2}(1-\frac{1}{3\zeta^2}   ) (R\Lambda)^{2}~.
\ee
Equivalently,
the scale parameters in the two schemes are related by
$\Lambda_{\rm ST} =\sqrt{3}\zeta\Lambda_{\rm PT}$, implying that
the unification scales in the two schemes are also related 
in the same way:
\be
M_{s} &=& \sqrt{3} \zeta M_{\rm GUT}^{\rm (PT)}
\simeq 0.748 M_{\rm GUT}^{\rm (PT)}~.
\ee

As a next task we would like to compute the Kaluza-Klein threshold
effect in the same way as in eq.~(\ref{evol4}).
We find that 
\be
\frac{8\pi^2}{g_{\rm ST}^2(\Lambda)}-
\frac{8\pi^2 }{g_{\rm ST}^2(1/R)}
& \simeq & -(b 
-\tilde{b})~ \ln (R\Lambda)
-\tilde{b} ~\left\{~\frac{\pi}{6\zeta^2}~[~
(R\Lambda)^{2}-1~]+\Delta_{\rm ST}~\right\}~,
\label{gst}
\ee
for $\Lambda > (1/R) e^T >1/R$, where (see also ref.~\cite{ross2})
\be
\Delta_{\rm ST} &=& T+
\frac{\pi}{6\zeta^2}(1-e^{2T})
+\int_{-\infty}^{T}dt ~\int_{-1/2}^{1/2}d\tau_1 
[~\vartheta_3^2 (i \sqrt{1-\tau_1^2}\zeta^2 e^{-2 t})-1~]\nn \\
& \simeq & 1.46 ~.
\label{delta71}
\ee
Since the string theory 
can be effectively regarded as  the PT
scheme with 
\be
r &=& 3\zeta^2~,
\ee
as we can see from eq.~(\ref{gst}),
we would like to compare the threshold
effect given in eq.~(\ref{delta71}) with that  in this
scheme. We find
\be
\Delta_{\rm PT}(r=3 \zeta^2)&=& 
T+
\frac{\pi}{6\zeta^2}(1-e^{2T})
\int_{-\infty}^{T}dt
\left[~
\vartheta_3^2 \left(\frac{i 3\zeta^2 e^{-2 t}}{\pi}\right)-1~
\right] \nn \\
& \simeq & 1.46~.
\ee
So we may conclude that the effective theory
with $r=3 \zeta^2$ can accurately describe the string theory 
for the case $M_s \gg 1/R$.

\section{Summary and discussion}


The higher dimensional theories, as a possible framework able in
principle to unify all interactions, have a very long history starting
from the work of Kaluza and Klein in the twenties and will certainly
continue in the next century. In particular during the last thirty years
there is a lot of theoretical interest in the various higher dimensional
schemes, while recently we have witnessed an increasing interest due to
the possibility that might be observable experimental consequences related
to some large compactification radii.
 Apart from the well known field theory limit of string theories 
(see {\em e.g.} ref.~\cite{string}),
there have been made  many attempts to consider Yang Mills Theories in
higher dimensions (see {\em e.g.} refs.~\cite{rieman}-\cite{duff}
and references therein) 
with most well known those of the
supergravity framework \cite{duff}. 
Maybe to the extend one is interested mainly
in the low energy properties in four dimensions of a gauge theory defined
in higher dimensions an elegant reduction scheme is the Coset Space
one \cite{reduction,sugra}.

Gauge theories in higher dimensions are (perturbatively)
non-renormalizable by power counting. However to the extend that a higher
dimensional theory is defined at weak coupling, it has been suggested in
ref.~\cite{kobayashi1} 
and in the present paper that the ERG provides us with a natural
framework to study also the quantum behavior of the theory. The ERG
approach based on the Wilson RG is indeed appropriate for dealing with
higher dimensional gauge theories since its formulation is independent of
dimensions and permits us to compute radiative corrections in a
meaningful fashion.

As we have emphasized, the scale parameter 
introduced there (see eq.~(\ref{wj1})) is
the {\em infrared} cutoff parameter and indicates the 
energy scale at which the effective theory is defined.
The RG equation (\ref{rge1}) describes the flow of the effective theory
as  $\Lambda$ varies, and it reduces in the
weak coupling  limit in the derivative expansion approximation 
to the evolution equation of the coupling.
We have also considered other schemes such as
the momentum subtraction (MOM) scheme and the 
proper time (PT) regularization scheme. In the
PT scheme the scale parameter $\Lambda$ 
is nothing but the ultraviolet
cutoff parameter.  Within this scheme it is indeed difficult
to understand that the coupling ``runs'' with $\Lambda$.
However, comparing all the schemes we have seen that
the ultraviolet cutoff parameter introduced in the PT
scheme and the subtraction scale introduced in the MOM scheme
are just the scale parameter of the ERG, and moreover,
that these regularization schemes give very similar results 
as far as the evolution of the coupling is concerned if one
carefully eliminates the apparent regularization scheme
dependence. We have arrived at the conclusion 
that the Kaluza-Klein threshold effect can be very small in 
certain regularization schemes; e.g.,
it is minimum for the $\pi$-scheme among the PT schemes.
This feature of the Kaluza-Klein thresholds originates from the fact
that, contrary to the renormalizable case, the lowest order 
$\beta$-function is regularization scheme dependent.

In a concrete application of our results we found that 
in fact the prediction of $\alpha_3(M_Z)$ resulting from 
gauge coupling unification in the MSSM with the Kaluza-Klein 
tower only in the gauge and Higgs 
supermultiplets \cite{dienes1} does not change practically,
in accord with the result of ref.~\cite{dienes1}.
We have also compared the Kaluza-Klein threshold effect
in a string theory and its effective field theory,
and we found that the string theory result
averages those of the field theory in different
regularizations, defining therefore an average 
regularization scheme of the effective theory.
Surprisingly, the Kaluza-Klein threshold effect
calculated in the effective theory with the average 
regularization can very well approximate
the corresponding string theory result.

\noindent
{\large \bf Acknowledgments}\\
Two of the authors (J.K and G.Z.) thank Alex Pomarol for discussions
and the organigers of the Summer Institute '99 at Yamanashi, which
gave the opportunity for extended discussions on the present work.

\end{document}